\documentclass[]{JHEP3}
\usepackage{epsfig}
\usepackage{cite}
\usepackage{amssymb}
\usepackage{rotating}

\title{Using Big Bang Nucleosynthesis to Extend CMB Probes of Neutrino Physics}

\author{M. Shimon$^{1}$, N.J. Miller$^{2}$, C.T. Kishimoto$^{3}$, C.J. Smith$^{4}$, 
G.M. Fuller$^{5}$ and B.G. Keating$^{6}$\\
$^{1,2,5,6}$Center for Astrophysics and Space Sciences, University of California, 
San Diego, La Jolla, CA, 92093\\
$^{3}$Department of Physics and Astronomy, University of California, Los Angeles, 
CA, 90095\\
$^{4}$Department of Physics, Arizona State University, Tempe, AZ, 85287\\
\\
$^{1}$meirs@mamacass.ucsd.edu, $^{2}$nmiller@physics.ucsd.edu, 
$^{3}$ckishimo@physics.ucsd.edu, $^{4}$christel.smith@asu.edu, 
$^{5}$gfuller@ucsd.edu, $^{6}$bkeating@ucsd.edu}

\abstract{
We present calculations showing that upcoming Cosmic Microwave Background (CMB) 
experiments will have the power to improve on current constraints on neutrino 
masses and provide new limits on neutrino degeneracy parameters. The latter could 
surpass those derived from Big Bang Nucleosynthesis (BBN) and the 
observationally-inferred primordial helium abundance. These conclusions derive from 
our Monte Carlo Markov Chain (MCMC) simulations which incorporate a full BBN nuclear 
reaction network. This provides a self-consistent treatment of the helium abundance, 
the baryon number, the three individual neutrino degeneracy parameters and other 
cosmological parameters. Our analysis focuses on the effects of gravitational 
lensing on CMB constraints on neutrino rest mass and degeneracy parameter. We find 
for the PLANCK experiment that total (summed) neutrino mass $M_{\nu} > 0.29$ eV could 
be ruled out at $2\sigma$ or better. Likewise neutrino degeneracy parameters 
$\xi_{\nu_{e}} > 0.11$ and $\vert \xi_{\nu_{\mu/\tau}} \vert > 0.49$ could be 
detected or ruled out at $2\sigma$ confidence, or better. For POLARBEAR we find that 
the corresponding detectable values are $M_\nu > 0.75\,{\rm eV}$, $\xi_{\nu_{e}} > 0.62$, 
and $\vert \xi_{\nu_{\mu/\tau}}\vert > 1.1$, while for EPIC we obtain 
$M_\nu > 0.20\,{\rm eV}$,  $\xi_{\nu_{e}} > 0.045$, and 
$\vert\xi_{\nu_{\mu/\tau}}\vert > 0.29$. 
Our forcast for EPIC demonstrates that CMB observations have the potential to set 
constraints on neutrino degeneracy parameters which are better than BBN-derived 
limits and an order of magnitude better than current WMAP-derived limits.}

\preprint{}


\begin{document}
                                                                                
\section{Introduction}

The CMB is a sensitive probe of basic cosmological 
parameters such as the spatial curvature of the 
universe and the energy density in baryons, dark 
matter, and dark energy.  Fundamental neutrino properties, 
such as their masses and the effective 
number of relativistic degrees of freedom, are 
already constrained by the CMB. Forecasts 
for future CMB experiments, e.g.\cite{lesgourgues.2006}, 
indicate that neutrino properties will be 
constrained with unprecedented accuracy.
These constraints, together with results 
from next-generation terrestrial experiments, 
may enable otherwise unobtainable insights 
into fundamental neutrino physics.
These results will be 
complementary to future terrestrial experiments 
as well as other cosmological probes 
({\it e.g.}, galaxy surveys
\cite{hu.1998, seljak.2006, mactavish.2006, tegmark.2006},
Ly$\alpha$ systems \cite{croft.1999, gratton.2008},
joint CMB and galaxy surveys 
\cite{tegmark.2004, hannestad.2003, elgaroy.2003, bernardis.2008}, 
weak lensing 
\cite{cooray.1999, abazajian.2003, hannestad.2006, kitching.2008} 
and joint CMB and weak lensing \cite{ichiki.2009, tereno.2009}).  

Most CMB features are imprinted at the epoch of recombination. 
However, post-recombination effects that introduce secondary 
temperature anisotropy ({\it e.g.}, lensing of the CMB by large 
scale structure (LSS) and the late integrated Sachs-Wolfe (ISW) 
effect) and polarization (CMB lensing by LSS) can be used to set 
tighter constraints on certain cosmological parameters.
Although neutrinos only weakly interact, they have been present 
for the entire history of the universe and can leave their imprint 
on both the CMB and LSS.  
This allows high-sensitivity and high-resolution CMB experiments to 
probe neutrino properties through the effect of the neutrinos on LSS.

The impact of neutrinos on the CMB strongly depends on their rest 
masses. Solar and atmospheric neutrino oscillation experiments 
have shown that at least two neutrino states are massive \cite{schwetz.2008}.
These neutrino experiments are sensitive to the differences in 
the squares of the neutrino masses, but not to their absolute 
mass scale (solar: 
$\delta m_{21}^{2}=7.65^{+0.23}_{-0.20}\times 10^{-5} ~{\rm eV}^{2}$; 
atmospheric: $\vert \delta m_{31}^{2} 
\vert =2.40^{+0.12}_{-0.11}\times 10^{-3}~{\rm eV}^{2}$).  
In addition, these neutrino experiments have not 
yet determined the sign of $\delta m_{31}^2$.  If it is positive, 
then the neutrino mass states are in the normal hierarchy, with 
two lighter mass states and one heavier mass state;  otherwise, 
if it is negative, then the neutrino mass states are in the 
inverted hierarchy with two heavier mass states and one lighter 
mass state. To pin down the three neutrino masses, a third 
independent measurement is required, for example, a measurement 
of the total, summed neutrino mass, $M_\nu \equiv \sum_{i=1,2,3} m_{\nu_i}$. 
Laboratory measurements of the neutrino mass-squared differences, 
imply that at least one neutrino mass must exceed $0.049\,{\rm eV}$. 
Thus, resolving the neutrino mass hierarchy ({\it i.e.,} mass ordering of 
the solar and atmospheric mass-squared doublets) may require sensitivity 
to the total neutrino mass of $M_\nu < 0.1\,{\rm eV}$.
For $M_\nu \gtrsim 0.1 ~{\rm eV}$, the two mass hierarchies 
are indistinguishable, but if the total neutrino mass could be 
constrained below this level, the inverted mass hierarchy would be ruled out. 

Another fundamental issue is how well cosmological 
probes can constrain the neutral lepton number.  
The lepton number residing in thermal neutrino seas can be 
characterized by neutrino degeneracy parameters, 
$\xi_{i} = \mu_{i}/ k_{B}T_\nu$ (where $\mu_{i}$ is the neutrino 
chemical potential of the $i$th species 
($\nu_{e}$, $\nu_{\mu}$ or $\nu_{\tau}$), 
$k_{B}$ is the Boltzmann constant and $T_\nu$ is the neutrino 
temperature) where neutrinos have a Fermi-Dirac distribution.
Current CMB data does not require the inclusion 
of neutrino chemical potentials in the cosmological model. 
In standard cosmology, the neutrino degeneracy 
parameters are assumed to be zero.  
However, there are a number of non-standard 
mechanisms that could lead to large neutral 
lepton asymmetries 
\cite{shi.1996, casas.1999, kawasaki.2002, 
yamaguchi.2003}.  Although calculations 
suggest that these asymmetries may equilibrate 
in the early universe \cite{dolgov.2002, 
savage.1991, Wong. 2002, abazajian.2002}, it is interesting 
to treat the lepton asymmetries in the three 
neutrino flavors independently.

Extracting neutrino masses from LSS 
tracers should account for the possibility that their 
chemical potentials do not vanish. A detection of nonvanishing 
neutral-lepton-asymmetry may have far-reaching implications.
The current best upper limits on neutrino degeneracy 
parameters, which are invariant under cosmological 
expansion, 
are provided by a comparison of big bang nucleosynthesis (BBN) 
calculations with the observed abundance of light elements, especially 
${}^4{\rm He}$ \cite{kneller.2001}. Current upper limits on $\xi$ 
from analysis of the  CMB are of order unity, while upper limits 
from BBN are on the order $\xi\sim 0.1$.  In this work we explore 
how these limits may be tightened by using future cosmological data.

This paper discusses the constraints on neutrino masses 
and degeneracy parameters that can be obtained from CMB 
data alone. In particular, we study the experimental capacity of PLANCK 
\footnote{http://www.rssd.esa.int/index.php?project=planck}, 
POLARBEAR
\footnote{http://bolo.berkeley.edu/polarbear/} and EPIC
\cite{bock.2009}
to constrain these parameters.
We constructed a joint BBN+CMB pipeline which 
self-consistently solves for the helium fraction, $Y_{p}$, given 
the other cosmological parameters and allows all three neutrino 
chemical potentials to vary independently of each other.
The helium fraction is not an independent parameter in our 
analysis (a similar approach was adopted in 
\cite{popa.2008, shiraishi.2009, hamann.2008}). 
Rather, we employ a BBN code \cite{wagoner.1969, wagoner.1973, kawano.1992} 
to self-consistently obtain $Y_{p}$ from a given set of other 
cosmological parameters, such as $\Omega_{b}$, $H_{0}$ 
and $\xi_{\nu_{e}}$, $\xi_{\nu_{\mu}}$ and $\xi_{\nu_{\tau}}$
Here the three neutrino degeneracy parameters are treated as phenomenological 
time-independent parameters, although models of time-dependent neutrino 
chemical-potentials have also been considered in the literature, e.g. \cite{yamaguchi.2003}.
Note, however, that
neutrino oscillations at the solar mass-squared splitting 
scale can ``even-up'' the lepton numbers for the different neutrino flavors - 
a process suggested by \cite{savage.1991} and shown to work more or less 
efficiently (to within a factor of ten) by \cite{abazajian.2002}, 
\cite{dolgov.2002} and \cite{Wong.2002}. 
The ultimate effect of this oscillation-driven process would be to keep all 
the lepton numbers all the same and subsequently fixed with time.
$Y_{p}$ is an important ingredient in the physics of 
recombination since it determines the Silk 
damping scale for a fixed baryon number. 
Earlier works discussing the implications of precise 
CMB observations on helium abundance inference are, e.g. 
\cite{trotta.2003, ichikawa.2006, ichikawa.2007} and more 
recently \cite{komatsu.2010}.
Our analysis also benefits from CMB lensing extraction 
achieved by employing the standard quadratic estimators 
of the lensing potential \cite{Huokamoto.2001}. 
This is important in exploring neutrino physics 
since 
it has been demonstrated that most 
of the information on neutrino parameters is 
encapsulated in CMB lensing 
\cite{kaplinghat.2003, lesgourgues.2006}. 

This work adds to previous efforts 
\cite{popa.2008, shiraishi.2009, hamann.2008}) which have attempted 
to constrain the neutrino degeneracy parameters from CMB or CMB+BBN 
by including gravitational lensing extraction of the CMB 
allowing the various degeneracy parameters to vary 
independently. Recently, a similar analysis for WMAP5 
was carried out which allowed 
$\xi_{\nu_{e}}\ne\xi_{\nu_{\mu}}=\xi_{\nu_{\tau}}$ 
\cite{shiraishi.2009}.
The PLANCK, POLARBEAR, and EPIC experiments have even higher 
sensitivity and resolution than WMAP. This can facilitate 
lensing extraction, allowing them to better probe neutrino 
parameters.

The paper is organized as follows. In section 2, we discuss 
the effects of neutrinos on BBN and the growth of structure. 
Section 3 describes our Monte Carlo Markov Chain (MCMC) simulation 
and the modifications we introduced in CAMB.
The degeneracies of neutrino mass and helium abundance 
with other parameters are especially relevant for parameter 
estimations from CMB observations and are therefore extensively 
discussed in section 4. We describe our results in section 5 
and conclude in section 6.

\section{Neutrinos and Neutral-Lepton Degeneracy}

\subsection{Definitions and Basic Quantities}

Over the history of the universe considered in this work the 
distribution functions of neutrinos ($\nu$) and 
anti-neutrinos ($\bar\nu$) with physical momentum $p$ are
\begin{eqnarray}
f_{\nu}(p;T_{\nu},\xi) & \approx & 
\frac{1}{e^{\frac{p}{T_{\nu}}-\xi}+1}\nonumber\\
f_{\bar{\nu}}(p;T_{\nu},\xi)& \approx & 
\frac{1}{e^{\frac{p}{T_{\nu}}+\xi}+1} ,
\label{eq:distributionfunction}
\end{eqnarray}
where $\xi\equiv\mu / T_\nu$ is the degeneracy parameter 
and $T_\nu$ is the time-dependent neutrino temperature \cite{fuller.2009}.
From here on, we will use natural units where 
$\hbar = c = k_{\rm B} = 1$. 
The degeneracy parameter is a comoving invariant. We assume that at some 
point in the early universe neutrinos and anti-neutrinos were in thermal and 
chemical equilibrium with the photon-baryon plasma and therefore 
$\xi_\nu + \xi_{\bar\nu} = 0$.
The cosmic neutrino background (C$\nu$B) temperature is inversely 
proportional to the cosmological scale factor, $a$, and is related 
to the (post-recombination) CMB blackbody temperature by 
$T_\nu = (4 / 11)^{1/3} T_{\rm CMB}$.

It is convenient to write the neutrino energy density and pressure in 
terms of the comoving momentum, $q = p a$ \cite{lesgourgues.1999}:
\begin{eqnarray}
\rho_\nu + \rho_{\bar\nu} & = & \frac{a^{-4}}{2 \pi^2} 
\int_0^\infty q^2 \,d q \sqrt{q^2 + (a M)^2} 
\left[ f_\nu (q / a; T_\nu, \xi ) + f_{\bar\nu} 
( q / a; T_\nu, \xi ) \right] \nonumber \\
P_\nu + P_{\bar\nu} & = & \frac{a^{-4}}{6 \pi^2} 
\int_0^\infty q^2 \,d q \frac{q^2}{\sqrt{q^2 + (a M)^2}} 
\left[ f_\nu (q / a; T_\nu, \xi ) + f_{\bar\nu} 
( q / a; T_\nu, \xi ) \right] ,
\end{eqnarray}
where $M\equiv m_\nu  / T_{\nu 0}$, with the C$\nu$B temperature 
at the current epoch, $T_{\nu 0} \approx 1.95 ~{\rm K}$.

The degeneracy parameter is related to the neutral lepton number,
\begin{equation}
L_\nu \equiv \frac{n_\nu - n_{\bar\nu}}{n_\gamma} = 
\frac{1}{33 \zeta (3)} \left( \pi^2 \xi + \xi^3 \right) ,
\end{equation}
where $n_\nu$, $n_{\bar\nu}$ and $n_\gamma$ are the number 
densities of neutrinos, anti-neutrinos and photons respectively, 
and $\zeta(3) \approx 1.202$ is the Riemann zeta function with argument 3.

The effective number of relativistic species, $N_{\rm eff}$, is a ratio 
between the energy density in a given relativistic species and the energy 
density of the same relativistic species with a thermal distribution and 
zero chemical potential.  If a neutrino and anti-neutrino of a given 
flavor both have Fermi-Dirac spectra and have equal and opposite 
degeneracy parameters [{\it e.g.}, Eq. (\ref{eq:distributionfunction})], 
then the effective number of relativistic species contributed by this 
neutrino flavor is 
\begin{equation}
N_{\rm eff} = 1+\frac{30}{7}\left(\frac{\xi}{\pi}\right)^{2}
+\frac{15}{7}\left(\frac{\xi}{\pi}\right)^{4}.
\end{equation}

The change in the total effective number of relativistic species is often 
used to describe the effects of non-standard neutrino distributions.  
When different degeneracy parameters are introduced for each neutrino 
flavor, the change in the overall effective number of relativistic species is
\begin{equation}
\Delta N_{\rm eff} = \sum_{i} \left[\frac{30}{7}
\left(\frac{\xi_{i}}{\pi}\right)^{2}+\frac{15}{7}
\left(\frac{\xi_{i}}{\pi}\right)^{4}\right].
\end{equation}
$\Delta N_{\rm eff}$ is a useful parameter when the detectable effects of 
neutrinos on the CMB depend on the contribution of these particles to the 
energy density in radiation.  However, upcoming CMB experiments will have 
the sensitivity to probe effects that are dependent on the distribution of 
neutrino energies.  In addition, BBN abundances are sensitive to neutrino 
energy distributions.  Hence, $\xi_\nu$ will be more useful than  
$\Delta N_{\rm eff}$ in the analysis of upcoming CMB experiments.

\subsection{Neutrino Effects on Cosmology}

\begin{figure*}
\includegraphics[width=0.8\linewidth]{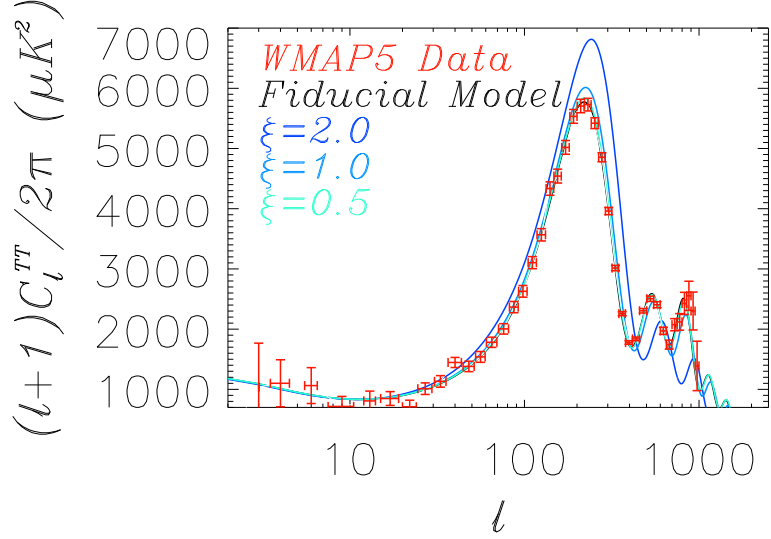}
\caption{The calculated CMB temperature anisotropy power spectrum 
for $\xi_\nu = 0$ (fiducial model), $0.5$, $1.0$, and $2.0$. 
The WMAP5 data points are included for reference. \label{fig:wmap5}}
\end{figure*}

Neutrinos have a wide range of effects on the evolution of the universe. 
In the early universe they participate in the reactions that determine 
the neutron-to-proton ratio which, in turn, affects the abundances of the 
light elements produced during BBN.
Later, at a redshift of $z\approx 3200$, the energy density in the 
C$\nu$B helps determine the epoch of matter-radiation equality.  
At recombination, $z\approx 1100$, the universe was not purely matter 
dominated, implying that gravitational potential wells had decayed, 
slightly.  This leads to the early ISW effect, which boosts the CMB 
temperature anisotropy angular power spectrum on multipole scales 
associated with the horizon scale, $l \lesssim 200$.  Neutrinos play 
a role in this process because the fraction of the total energy density 
in the form of radiation (which is sensitive to neutrino masses and 
degeneracy parameters, Eq.\ 2.2) determines the amplitude of the ISW effect.  
This is the only effect of neutrino mass and degeneracy parameter that 
can be probed by WMAP and other moderate angular resolution experiments.  
Fig.\ \ref{fig:wmap5} shows the calculated CMB power spectrum for 
various $\xi$, along with the data points from WMAP5.  It is clear that 
$\xi > 1$ is excluded at $1 \sigma$ (assuming all other parameters 
are fixed). A global parameter analysis reaches a similar conclusion 
\cite{shiraishi.2009}.

An aspect highlighted in this paper is that stringent constraints on 
neutrino mass and degeneracy parameters can come from an analysis of 
CMB lensing.  A neutrino that is non-relativistic today could have 
been relativistic at higher redshifts. Non-relativistic neutrinos 
could be captured into potential wells, while relativistic neutrinos 
would act as hot dark matter (HDM) and would freely stream, resulting 
in an apparent suppression of structure formation during the epochs 
when the neutrinos are relativistic.  Precise measurements of the LSS 
power spectrum can be used to place constraints on neutrino masses 
and degeneracy parameters.

\subsubsection{BBN and Light Element Abundances}

BBN occurs at temperatures much higher than the current upper bounds on 
neutrino masses and therefore BBN calculations cannot constrain neutrino 
masses.  However, the neutrino degeneracy parameters impact BBN 
abundance-yields by affecting both the reaction rates that determine the 
neutron-to-proton ratio and the expansion-rate of the universe, which helps to 
determine how neutron-to-proton inter-conversion works.
The weak reactions that set the neutron-to-proton ratio are
\begin{eqnarray}
& \nu_e + n \rightleftharpoons p + e^- \nonumber \\
& \bar\nu_e + p \rightleftharpoons n + e^+ \\
& n \rightleftharpoons p + e^- + \bar\nu_e ~. \nonumber
\label{eq:weakrxn}
\end{eqnarray}
The rates of these reactions depend on the number density and energy 
spectrum of $\nu_e$ and $\bar\nu_e$, which in turn depend on the electron 
neutrino degeneracy parameter, $\xi_{\nu_e}$ 
\cite{wagoner.1967, smith.2009, abazajian.2005, smith.2006, kneller.2001}.
These reaction rates compete with the expansion rate of the universe 
which is determined by the total energy density; the latter also depends 
on all neutrino degeneracy parameters, Eq.\ (2.2). It is clear, therefore, 
that BBN distinguishes $\xi_{\nu_{e}}$ from 
$\xi_{\nu_{\mu}}$ and $\xi_{\nu_{\tau}}$, making for a nontrivial interplay 
between the neutrino degeneracy parameters and the light element abundances, 
particularly ${}^4{\rm He}$.

Combined analysis of the CMB (BOOMERANG and DASI experiments), 
BBN (helium and deuterium abundance) and SNIa data yield the following 
$2\sigma$ limits \cite{hansen.2001} 
\begin{eqnarray}
 -0.01<\xi_{\nu_{e}}<0.22\nonumber\\
\vert \xi_{\nu_{\mu},\nu_{\tau}} \vert <2.6.
\end{eqnarray}
If oscillation between the three neutrino species results in equilibration 
of the asymmetries among the neutrino flavors
\cite{dolgov.2002, Wong.2002, savage.1991}, 
then the more stringent ${}^4{\rm He}$ constraint on $\xi_{\nu_e}$ 
applies to all neutrino flavors and BBN considerations suggest
\cite{kneller.2001, abazajian.2005, barger.2003}
\begin{equation}
\vert \xi_{\nu} \vert \lesssim 0.1.
\end{equation}
However, non-standard physics could lead to 
different degeneracy parameters for the three different neutrino flavors.

BBN determines the abundance of light elements, including the helium fraction, 
$Y_p$.  These abundances can be sensitive to the baryon closure fraction, 
$\Omega_b$, and the three neutrino degeneracy parameters.  In particular, 
$Y_p$ is determined principally by the neutron-to-proton ratio at temperatures 
$T \sim 100 ~{\rm keV}$.  This ratio is set by the competition between the 
weak reactions in Eq.\ (2.6).  As a result, $Y_p$ depends strongly on 
the neutrino degeneracy parameters.  

Until recently, the helium fraction was usually considered as a free 
parameter in CMB analyses.  Recent work 
\cite{hamann.2008, popa.2008, shiraishi.2009} has attempted to 
self-consistently include $Y_p$ as a non-independent parameter in 
CMB power spectra calculations.  It was noted in \cite{hamann.2008} 
that certain cosmological parameters are significantly biased when 
$Y_p$ is fixed at $Y_p = 0.24$ and consistency with BBN is ignored. 

Helium recombination occurs prior to hydrogen recombination. Therefore, 
for a fixed baryon closure fraction, the number density of free electrons 
at hydrogen recombination is a function of the helium abundance.  
The Silk damping scale is the scale over which temperature anisotropy 
and polarization will be washed out by free-streaming of photons between the onset 
and the end of decoupling.  This scale depends on the photon mean free path 
which is inversely proportional to the number density of free electrons.  
Increasing $Y_p$ reduces the number density of free electrons at hydrogen 
recombination, which increases the mean free path of the CMB photons.  
The result would be a suppression of correlations on larger angular
scales, which would shift Silk damping to lower multipole numbers.

\subsubsection{The Growth of Large Scale Structure}

While weak constraints on neutrino masses can be extracted from the 
primary CMB power spectra, adding probes of structure formation has 
the potential to significantly tighten these bounds.  Using CMB 
lensing rather than resorting to other cosmological probes of 
structure formation is nearly systematic-free, providing high 
fidelity constraints.

CMB lensing is a sensitive probe of any cosmological parameter 
that impacts the growth rate of gravitational potential wells. 
Current CMB data, combined with observational data from Type Ia 
Supernovae (SNIa) and baryon acoustic oscillations (BAO), 
constrains the total neutrino mass to the sub-eV level \cite{komatsu.2009}.
Since the lensed CMB is the result of  the integrated effect of 
the lensing of the primary CMB by structure formation, and the 
relevant redshift range for structure formation may overlap with 
the epoch where neutrinos transition from being relativistic to 
non-relativistic, the CMB can be a powerful tracer of neutrino 
masses and degeneracy parameters.  Additional leverage on 
neutrino free streaming comes from e.g., galaxy correlations, 
Ly$\alpha$ forest power spectra \cite{seljak.2006,goobar.2006} 
and weak galaxy lensing \cite{ichiki.2009}.

Tracers of the matter power spectrum, such as CMB lensing, are 
sensitive to the epoch when neutrino momenta were redshifted 
to a point where they are non-relativistic.  This is because 
non-relativistic neutrinos behave as a cold dark matter (CDM) 
and contribute to the growth of structure, while relativistic 
neutrinos behave as HDM and suppress structure on scales 
below their free streaming scale.  Thus, the epoch when neutrinos 
become non-relativistic is important in discerning the effect of 
neutrinos on large scale structure.  Both the neutrino mass and 
degeneracy parameter determine when neutrinos become non-relativistic.

Probes of the growth of structure in the universe indicate that 
CDM, rather than HDM, is the dominant 
component of matter. Neutrino masses of $0.2 - 0.3 ~{\rm eV}$ 
(consistent with the upper limits in the current neutrino mass 
constraints) are mildly relativistic at recombination which would 
result in the slight decay of gravitational potential wells at 
last scattering, leading to a primary ISW effect. 
For a spatially flat universe, $\Omega_{k}=0$, 
and a fixed dark energy density fraction, $\Omega_{\Lambda}$, changing 
the neutrino masses will change the amount of HDM at a given redshift 
at the expense of CDM. This will cause a relative suppression of 
structure formation at high redshifts. 
In turn, this will be reflected in the level of CMB lensing by LSS. 
Several forecasts for PLANCK, CMBPOL \footnote{http://cmbpol.uchicago.edu/} 
and other CMB experiments suggest that constraints on neutrino masses 
can be improved by a factor of three to four \cite{lesgourgues.2006}, 
provided the experiments have sufficiently high sensitivity and angular 
resolution to allow lensing extraction. As already mentioned, 
the CMB, galaxy redshift surveys, cluster abundances, 
Ly$\alpha$ and other sensitive probes of the growth of structure 
on scales of a few tens of Mpc can be employed to set sub-eV constraints 
on the total neutrino mass.  An intriguing question is whether 
these probes could distinguish between the normal and inverted 
neutrino mass hierarchies. Constraining the total neutrino mass below 
the required $\sim 0.1 ~{\rm eV}$ scale is a challenging task in the 
presence of astrophysical foregrounds and other systematics.  It was 
recently shown, e.g. \cite{seljak.2006, kitching.2008}, 
that by combining several cosmological probes, this (or similar) 
limit can be achieved.
However, it is important to be mindful of the assumptions that are made in 
achieving these limits and to what extent the systematics can be controlled.

The free streaming scale can be estimated as the proper distance traveled 
by a neutrino over the age of the universe. This gives an estimate of the 
neutrino free streaming scale, $\lambda_{\rm FS}$.  This scale is
\begin{equation}
\lambda_{\rm FS} = \left\langle \int_0^{t_0} v (t) \frac{d t}{a(t)} \right\rangle ,
\end{equation}
where $v(t)$ is the neutrino velocity, which decreases as the universe 
expands, $a(t)$ is the scale factor, $\langle \cdots\rangle$ denotes an 
average with respect to the neutrino energy distribution function, 
Eq.\ (\ref{eq:distributionfunction}), and $t_{0}$ is the time today.  
Matter overdensity on scales smaller than $\lambda_{\rm FS}$ will be 
suppressed by neutrino free streaming.  This suppression factor will 
be proportional to $\Omega_\nu$, the neutrino energy density in closure 
density units.

\begin{figure*}
\includegraphics[width=0.35\linewidth, angle=270]{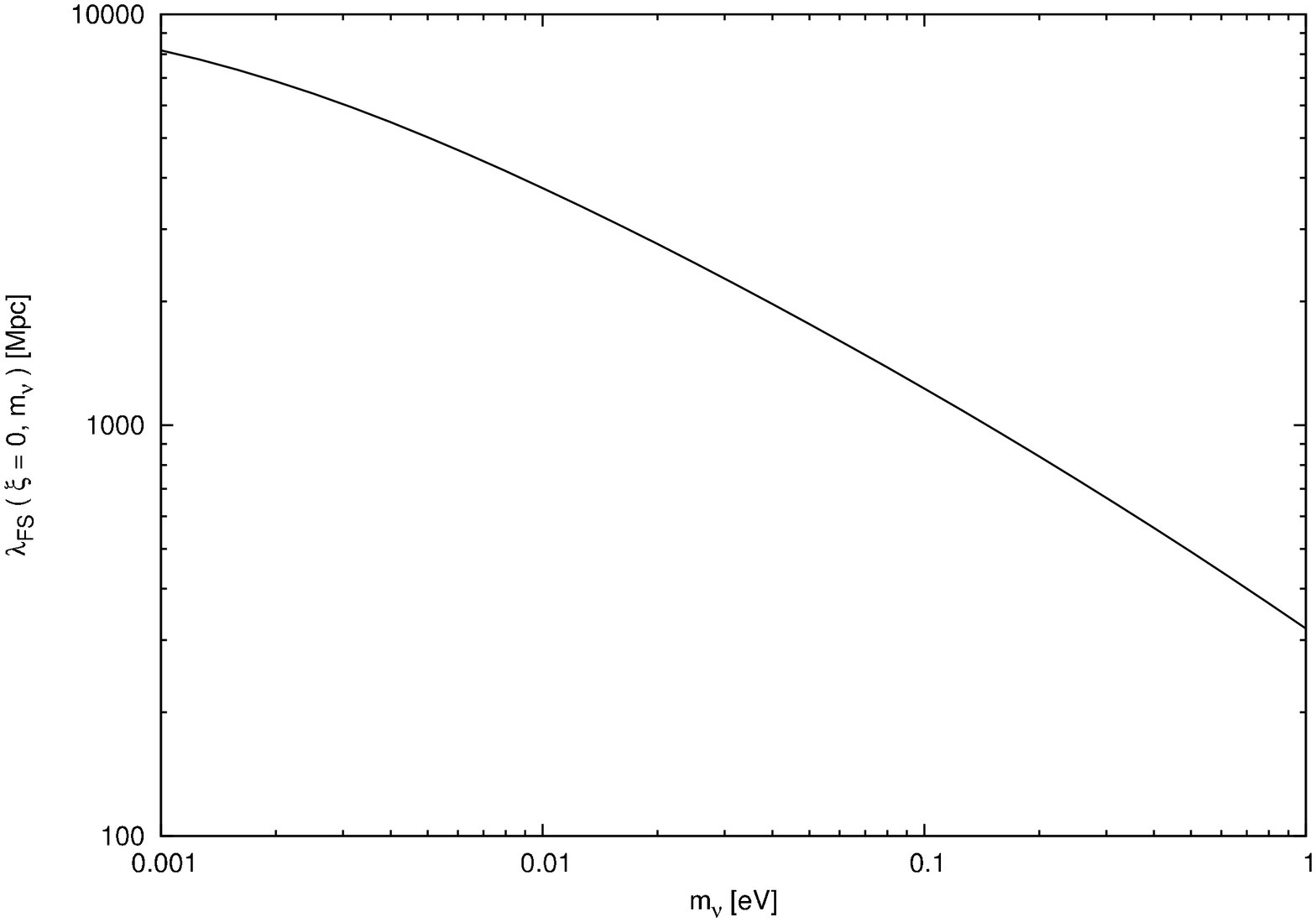}
\includegraphics[width=0.35\linewidth, angle=270]{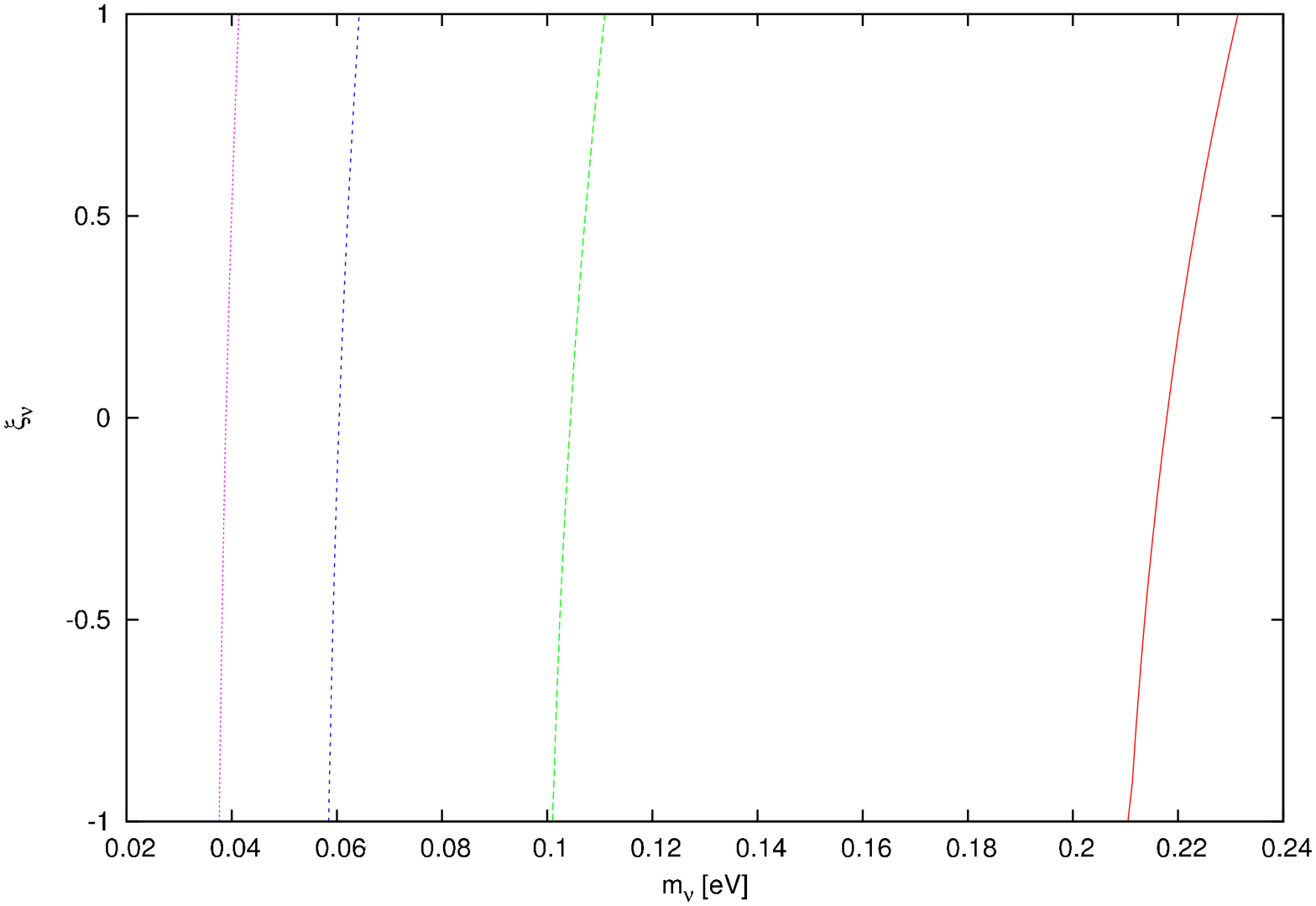}
\caption{Neutrino free streaming scale:  
The figure on the left is the neutrino free streaming scale as a 
function of neutrino mass with $\xi_\nu = 0$.  The figure on the 
right is a contour plot of constant free streaming scale in the 
$m_{\nu}$-$\xi$ plane; the contours from right to left correspond 
to $\lambda_{\rm FS} = 0.8,$ $1.2$, $1.6$, and $2.0~{\rm Gpc/h}$. 
\label{fig:freestreaming}}
\end{figure*}

The average free streaming scale for a neutrino species with mass $m_\nu$ 
and degeneracy parameter $\xi$ is 
\begin{eqnarray}
\lambda_{\rm FS} ( \xi, M ) & = & \frac{1}{F_2 (\xi)} \int_{q=0}^\infty 
\int_{t=0}^{t_0} \frac{q^2}{e^{q-\xi}+1} 
\frac{q \,d q}{\sqrt{q^2 + [a (t) M]^2}} \frac{d t}{a(t)}
\end{eqnarray}
where $M = m_\nu / T_{\nu 0}$
and $F_2 (\xi)$ is the Fermi integral of order two,
\begin{equation}
F_2 (\xi) \equiv \int_0^\infty \frac{q^2 \,d q}{e^{q-\xi}+1} .
\end{equation}
The free streaming scale dependence on
neutrino masses and degeneracy parameter is illustrated in Figure 
\ref{fig:freestreaming}. Note that the free streaming scale increases 
with decreasing mass and increasing degeneracy parameter.  
Care should be taken when simultaneously discussing neutrino degeneracy 
parameters (which are related to flavor states) and neutrino masses 
(which are related to mass states) \cite{fuller.2009}.

\begin{figure*}
\includegraphics[width=0.5\linewidth]{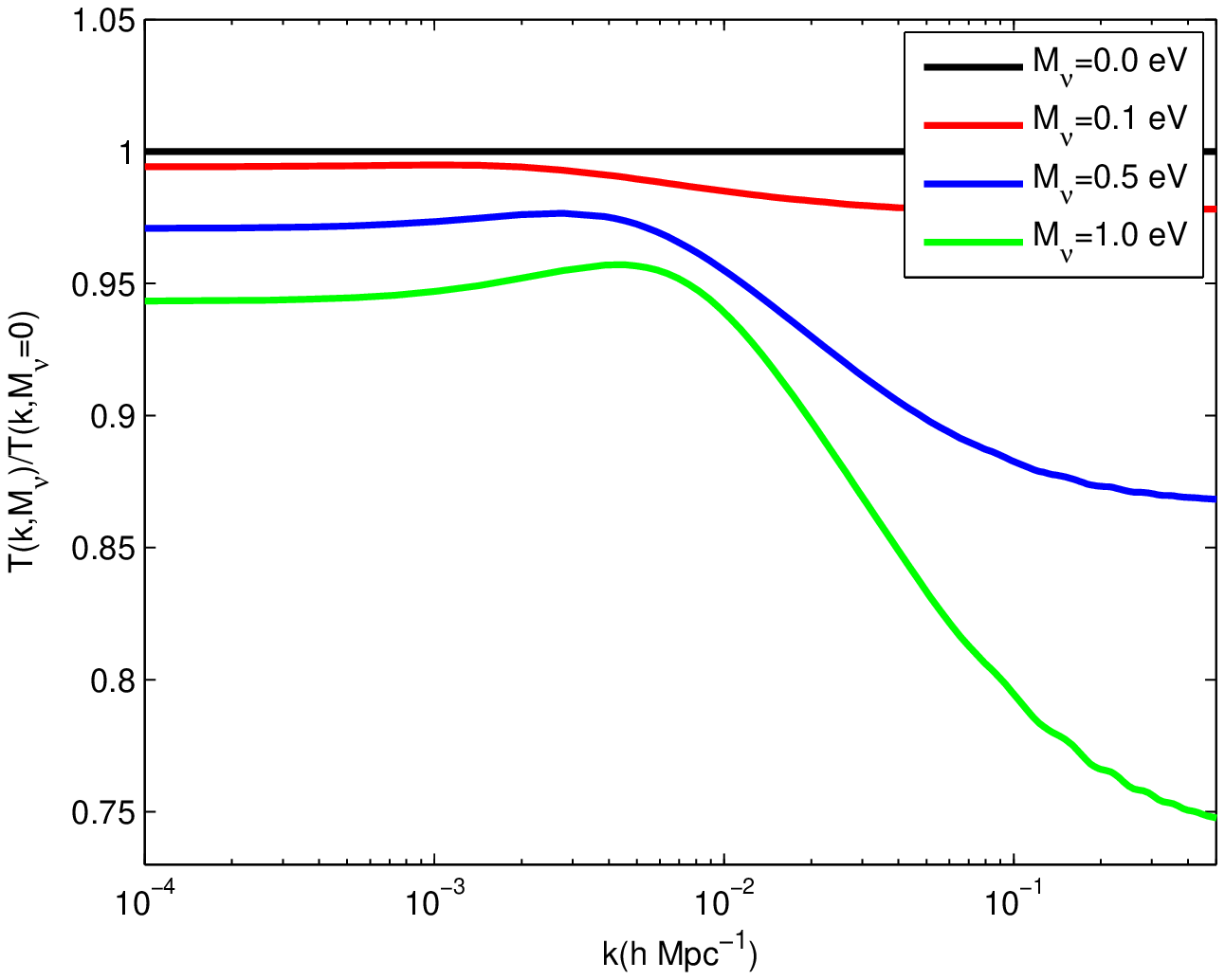}
\includegraphics[width=0.6\linewidth]{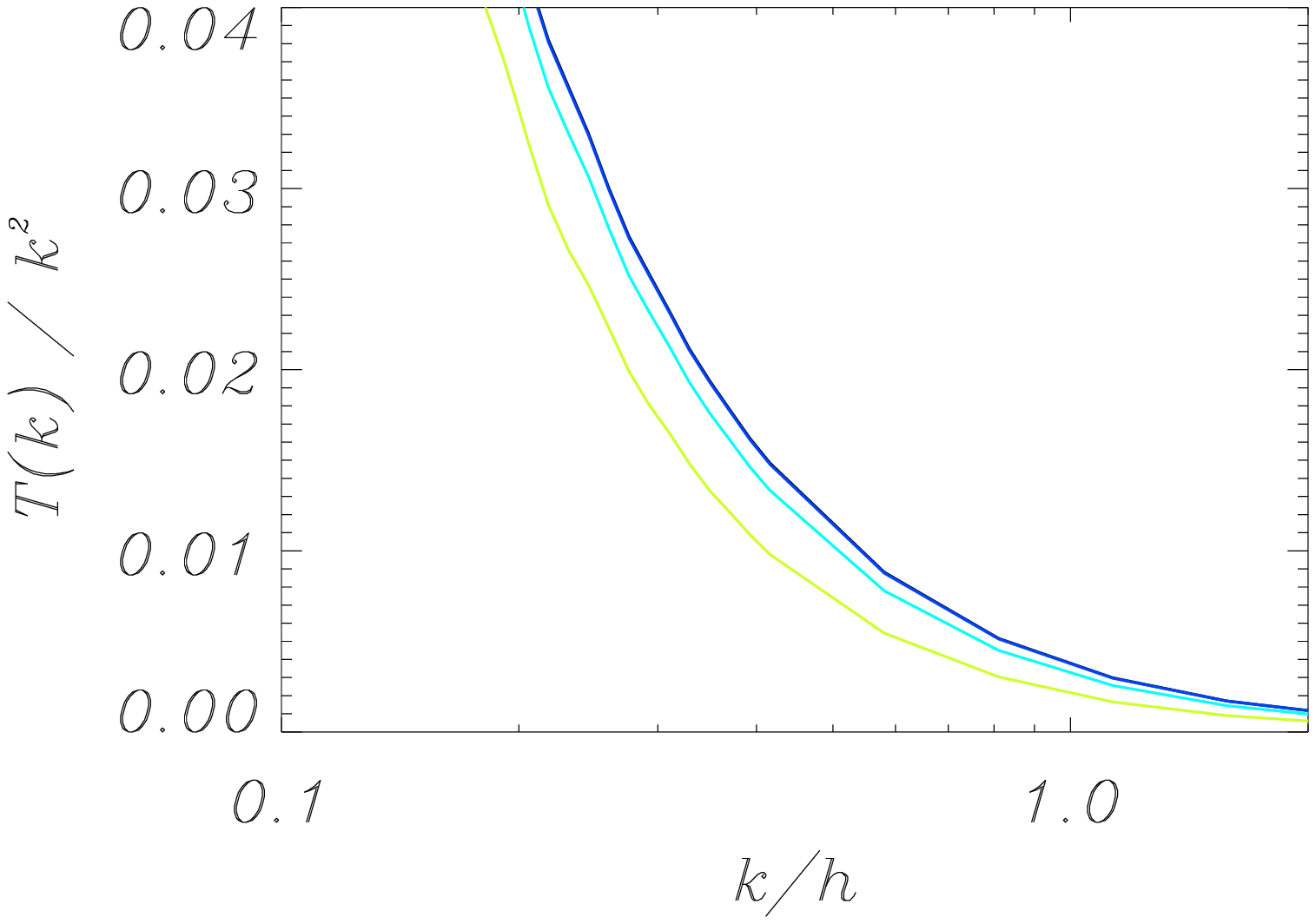}
\caption{Susceptibility of the transfer function to $M_{\nu}$ 
(left) and $\xi_{\nu}$ (right). 
The values used for $\xi$ are  0.1 (blue), 0.5 (cyan) and 1.0 
(yellow). \label{fig:transferfunction}}
\end{figure*}

Figure \ref{fig:transferfunction} illustrates how neutrino mass and 
degeneracy parameters affect the transfer function.  
The transfer function represents the effect of all physical processes 
that cause the primordial power spectrum to evolve into the matter 
power spectrum at latter epochs. The relation between the power 
spectrum $P_m (k;z)$ and the transfer function $T(k;z)$ can be 
written as
\begin{equation}
P_m (k;z) = A_{s} k^{n_s} T^2 (k;z) ,
\end{equation}
where $n_s$ and $A_{s}$ are the tilt and normalization of the primordial 
power spectrum. The suppression of the matter 
power spectrum on scales smaller than the neutrino free streaming scale 
is related to observable quantities. This could be obtained from 
galaxy surveys or inferred from the CMB angular power spectrum by  
deconvolving the lensing power spectrum.  The change in the matter power spectrum 
resulting from neutrino free streaming is \cite{hu.1997} 
\begin{equation}
\frac{\Delta P_m (k)}{P_m (k)} \approx - 8 \frac{\Omega_\nu}{\Omega_m} .
\end{equation}
Here, $P_m (k) \equiv P_m(k; z = 0)$.
The effect of non-vanishing neutrino mass is shown on the left side of 
Figure \ref{fig:transferfunction}.  As the neutrino mass increases, the 
free streaming scale decreases, leading to suppression of the transfer 
function at larger wavenumbers (note that all curves are normalized to 
the case of $m_{\nu}=0$ at low $k$).  The suppression of the transfer 
function at these large wavenumbers is more pronounced for larger 
neutrino masses because in this case neutrinos constitute a larger 
fraction of the dark matter.  The effect of non-zero neutrino degeneracy 
parameter is shown on the right side of Figure \ref{fig:transferfunction}. 
As the degeneracy parameter increases, the free streaming scale increases, 
leading to suppression of the transfer function at progressively smaller 
wavenumbers. 

\begin{figure*}
\includegraphics[width=0.5\linewidth]{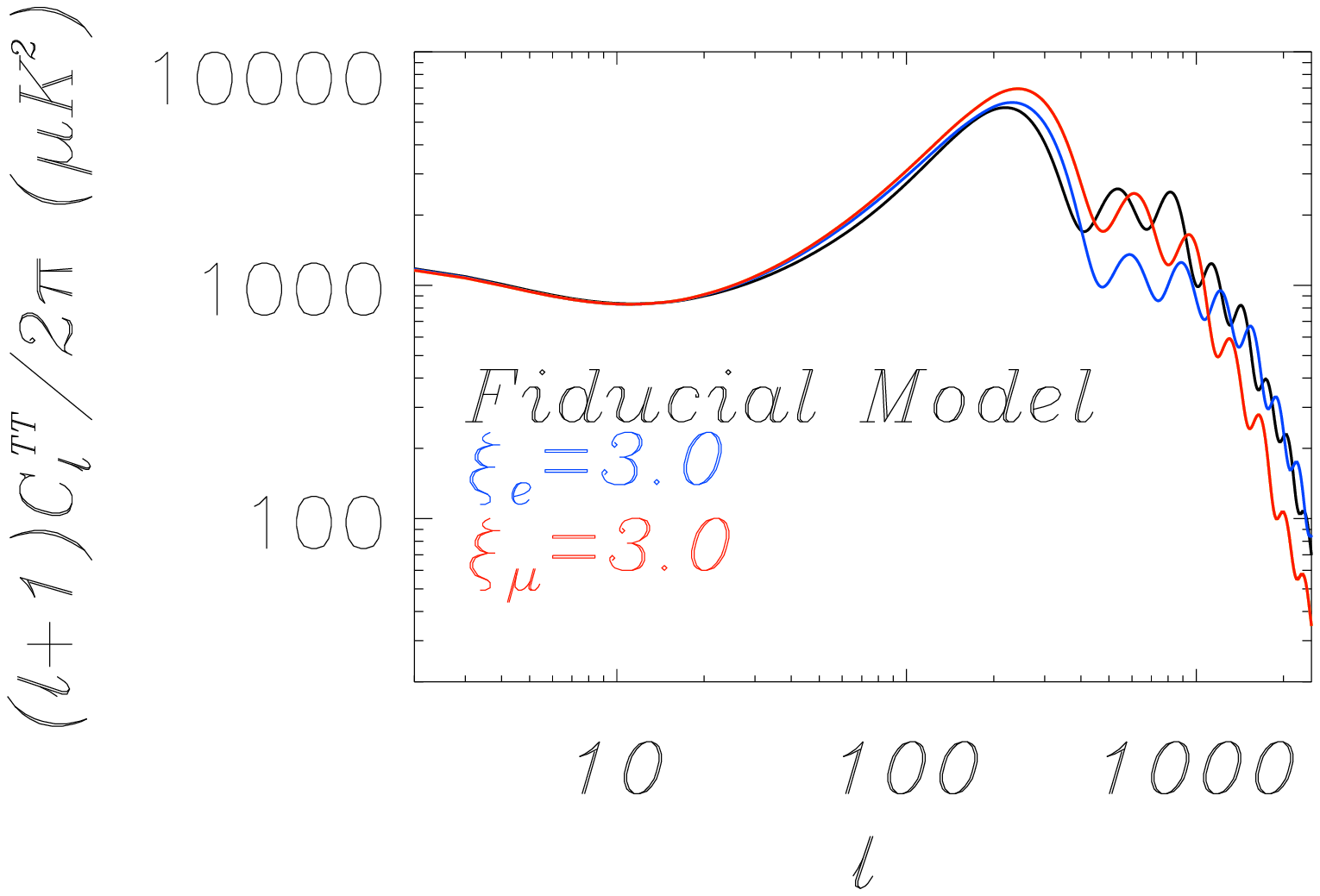}
\includegraphics[width=0.5\linewidth]{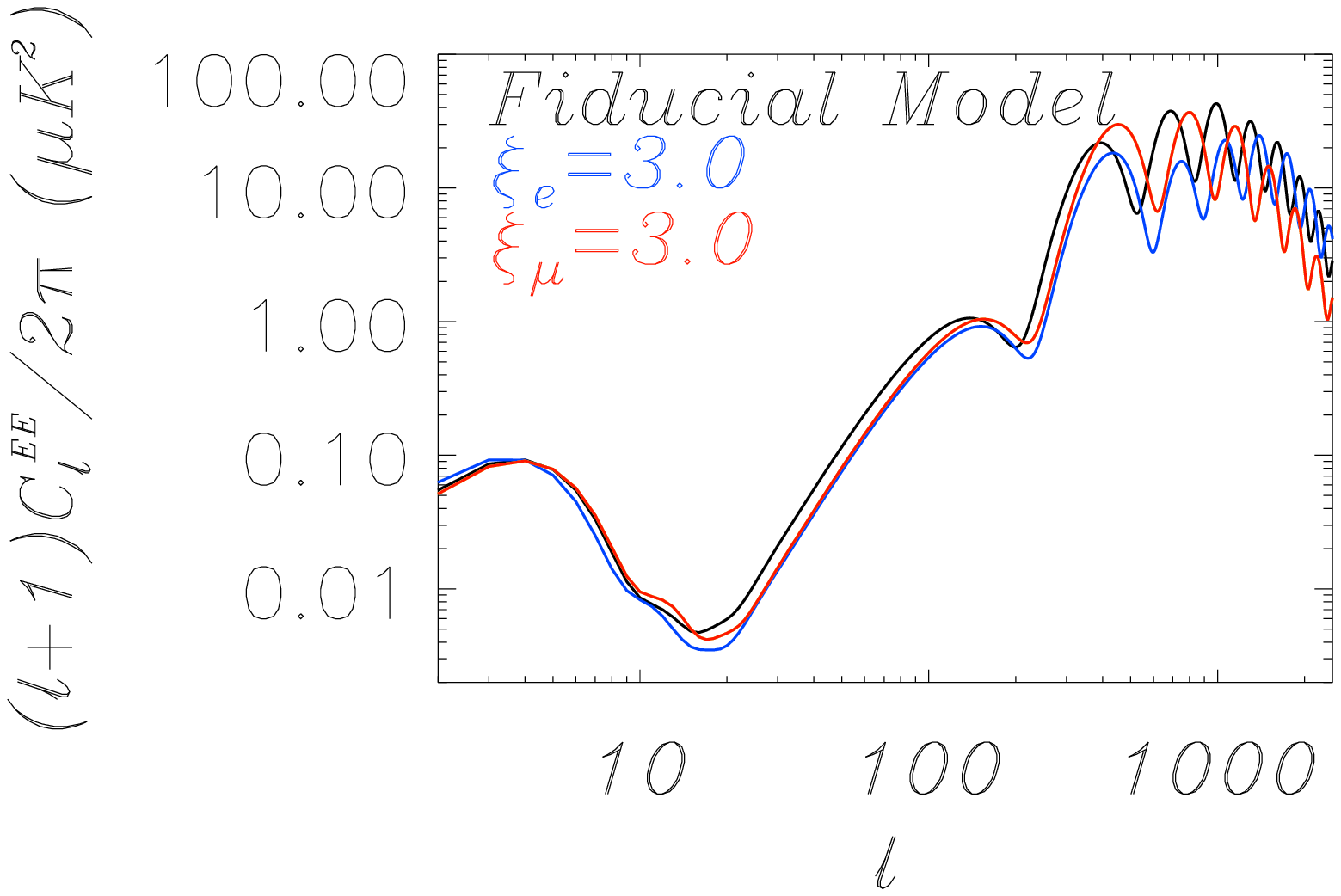}
\includegraphics[width=0.5\linewidth]{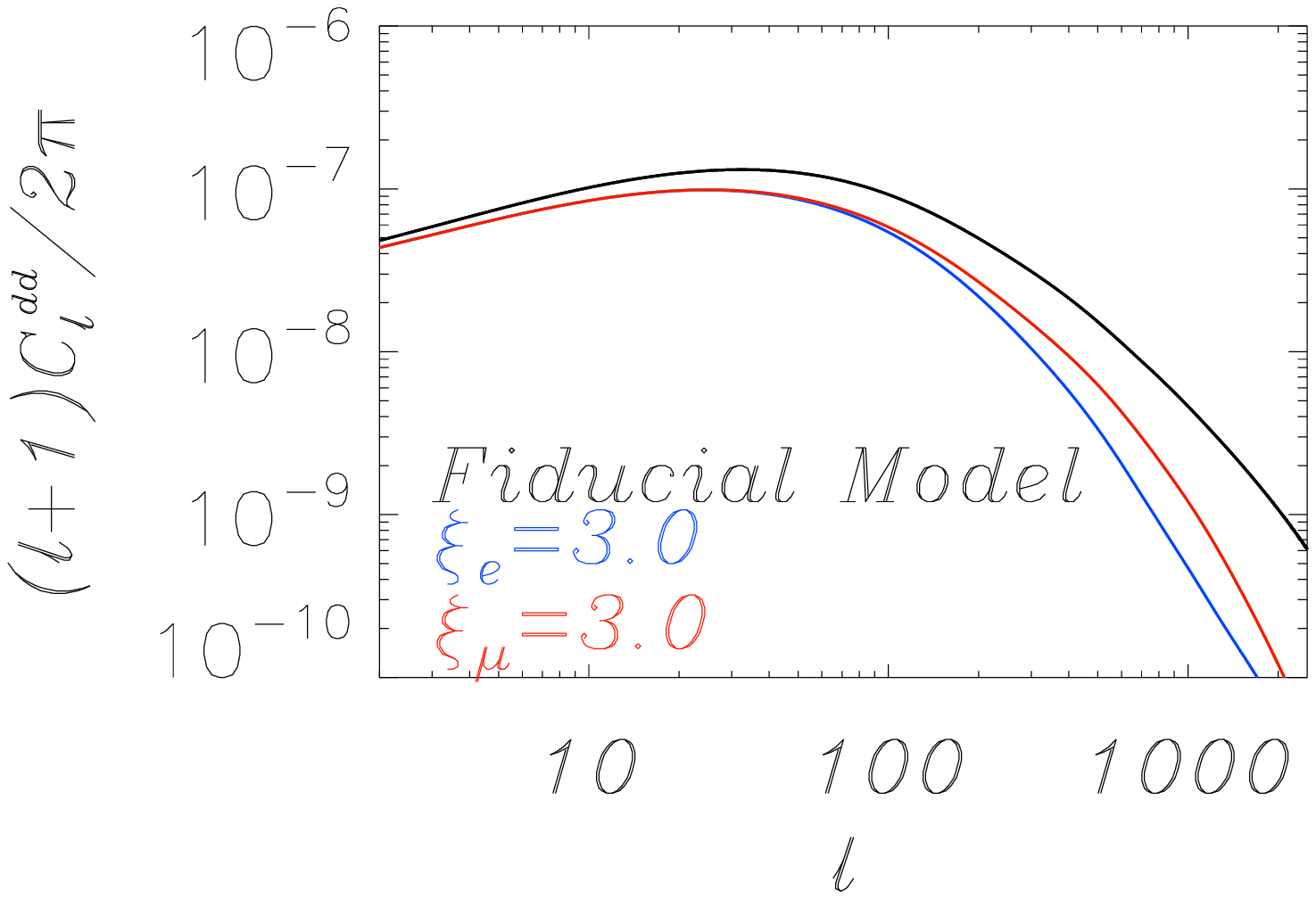}
\caption{CMB power spectra response to changing $\xi_\nu$: 
$C_{l}^{TT}$ (top-left), $C_{l}^{EE}$ (top-right), 
$C_{l}^{dd}$ (bottom). In all three plots the black 
curves correspond to the fiducial model 
($\xi_{\nu_{e}}=\xi_{\nu_{\mu}}=\xi_{\nu_{\tau}}=0$), 
the red curves correspond to a non-zero $\xi_{\nu_e}$ 
model ($\xi_{\nu_{e}}=3$, $\xi_{\nu_{\mu}}=\xi_{\nu_{\tau}}=0$), 
and the blue curves correspond to a non-zero $\xi_{\nu_{\mu,\tau}}$ 
model ($\xi_{\nu_{e}}=0$, $\xi_{\nu_{\mu}}=\xi_{\nu_{\tau}}=3$). 
 \label{fig:powerspectra}}
\end{figure*}

Figure \ref{fig:powerspectra} illustrates the effect of non-zero degeneracy 
parameters on the CMB temperature, polarization, and deflection angle power 
spectra (the latter is essentially a measure of the rms lensing deflection 
angle of the LSS, relevant to CMB lensing).  The most significant differences 
are at large multipoles, to which current CMB experiments are blind, but PLANCK, 
POLARBEAR and EPIC will be sensitive. Although degeneracy parameters 
$\xi_{\nu}=3$ are already ruled out by BBN and CMB data, we show these cases for 
illustrative purposes. From the plots of $C_{l}^{TT}$ and $C_{l}^{EE}$ we 
can see the effect of neutrino degeneracy parameters on scales from the acoustic 
horizon at recombination down to Silk damping scales. 
The power spectrum $C_{l}^{dd}$ for lensing deflection 
angle, $d$, is suppressed in the presence of nonvanishing $\xi_{\nu}$ at high $l$. 
This reflects the relative suppression in the transfer function at large wavenumbers. 
This effect, leads to suppression of the lensing-induced B-mode polarization that 
results from E-B conversion via CMB lensing by LSS. Therefore, the presence of 
a large $\xi_\nu$ can be constrained with high sensitivity CMB experiments 
which are capable of discerning small variations in the weak B-mode polarization.  

\section{CMB Code and Monte Carlo Simulation}

As in \cite{hamann.2008, popa.2008, lesgourgues.1999}, we modified 
the Boltzmann code CAMB \cite{camb} by replacing the neutrino distribution 
function (here $q\equiv p/T$)
\begin{eqnarray}
f_{\nu}(q)=\frac{1}{e^{q}+1}
\end{eqnarray}
with
\begin{eqnarray}
f_{\nu}(q;\xi)=\frac{1}{2}
\left(\frac{1}{e^{q+\xi}+1}+\frac{1}{e^{q-\xi}+1}\right)
\end{eqnarray}
everywhere, including in the expressions for energy 
density and pressure, as well as in the Liouville 
equation for neutrino density perturbations. 
We allowed the individual neutrino flavors to have 
three different degeneracy parameters 
$\xi_{\nu_e}$, $\xi_{\nu_\mu}$ and $\xi_{\nu_\tau}$. 

Neutrino masses are subject to experimental constraints from terrestrial neutrino 
experiments.  The neutrino masses and the degeneracy parameters are degenerate with 
various other cosmological parameters, presenting a challenge in attempting to 
determine these parameters using the CMB. For a given value of $\Omega_{\nu}$, 
both $m_{\nu}$ and $\xi_{\nu}$ are degenerate; for a fixed $\Omega_{\nu}$, 
increasing $\xi_\nu$ must be compensated by decreasing neutrino masses. To avoid 
this degeneracy in interpretation of our simulation results we consider
$m_{1}$, $m_{2}$, $m_{3}$, $\xi_{\nu_{e}}$, $\xi_{\nu_{\mu}}$ and 
$\xi_{\nu_{\tau}}$ as our basic parameters (in addition to the standard cosmological 
parameters). The three neutrino masses are constrained by the measured mass squared 
differences.  In this work, we used conservative gaussian priors for these differences:  
$\delta m_{21}^2 = 8.0 \pm 0.6 \times 10^{-5} ~{\rm eV}^2$ and 
$\delta m_{31}^2 = 2.4 \pm 0.6 \times 10^{-3} ~{\rm eV}^2$, which are consistent with 
the experimental values. In practice, these and projected future laboratory 
improvements in neutrino mass-squared uncertainties, have little effect on our 
analysis. The dominant uncertainties come from CMB data uncertainties. 

CMB data is encapsulated in the angular power spectra down 
to scales determined by the angular resolution of the 
specific experiment and its instrumental noise level 
as compared to the CMB signal. The instrumental noise $N_{l}$ 
in measuring the angular power spectrum for multipole $l$, for 
the autocorrelation of the temperature and polarization of the 
E- and B-modes are related for bolometric radiometers by 
$2 N_{l}^{TT} = N_{l}^{EE} = N_{l}^{BB}$.
The instrumental noise is uncorrelated between $T$, $E$, and $B$ 
and increases exponentially with multipole number,
\begin{eqnarray}
N_{l,\nu}^{ab}
=\delta_{ab}(\theta_{a}\Delta_{a})^{2}
\exp[l(l+1)\theta_{a}^{2}/8\ln 2] ,
\end{eqnarray}
where $a$ and $b$ are either $T$, $E$, or $B$.
Here, the noise at the frequency band centered 
at $\nu$ is a function of the corresponding 
beamwidth, $\theta_{a}$, and the noise per pixel in 
equivalent temperature 
units, $\Delta_{a}$.
To obtain the effective noise power contributed 
by all frequency bands in the experiment 
one adds them as if they were uncorrelated 
gaussians
\begin{eqnarray}
N_{l}^{aa}=[\sum_{\nu}(N_{l,\nu}^{aa})^{-1}]^{-1}.
\end{eqnarray}

We simulated parameter extraction from PLANCK, 
POLARBEAR and EPIC.
The sensitivity and resolution we considered for these experiments
are given in Table \ref{tab:specs}.
The CMB power spectra $C^{TT}$, $C^{TE}$, 
$C^{EE}$ and $C^{BB}$, together with the power spectrum 
of the deflection angle $C^{dd}$, and its 
cross-correlation with the temperature 
anisotropy, $C^{Td}$ are calculated by CAMB for a given fiducial 
cosmological model. When the $C_{l}^{Td}$ power 
spectra are calculated with the parameter \textbf{accuracy\_level=1},
there are very large oscillations at $l \approx 200$. These go away 
when the accuracy level is increased. In order to fix this in our simulation, 
the $C_{l}^{Td}$ at $l>200$ in the simulated data are replaced with values 
calculated when setting \textbf{accuracy\_level=5} in CAMB. When calculating 
the likelihood in CAMB, the proposed $C_{l}^{Td}$ are set to the same value 
as the ``experimental'' $C_{l}^{Td}$. We found that doing so does not affect 
the parameter uncertainties extracted from the MCMC simulation. 
All power spectra are assumed gaussian and are taken to be unlensed following 
the conclusion of \cite{lesgourgues.2006}. The noise in lensing reconstruction, 
$N_{l}^{dd}$, is a function of the observed power spectra (all four lensed 
$\tilde{C}_{l}$ with instrumental noise, Eq.(3.3), included) 
and the unlensed power spectra \cite{kaplinghat.2003} (without lensing, 
as obtained from CAMB for a fiducial cosmological model). In calculating 
$N_{l}^{dd}$ we employ the publically available code 
\cite{perotto.2006} which makes use of the quadratic estimators 
\cite{Huokamoto.2001}.

\section{$M_{\nu}$ and $\xi_\nu$ and Their Degeneracies with Other Parameters}

Neutrino masses and chemical potentials are both 
degenerate with each other and with other 
cosmological parameters.
We now discuss the main degeneracies of neutrino parameters
with other cosmological parameters.
In particular, we discuss the degeneracy of neutrino 
masses in section 4.1 and chemical potential 
in section 4.2.

\begin{figure*}
\includegraphics[width=0.5\linewidth]{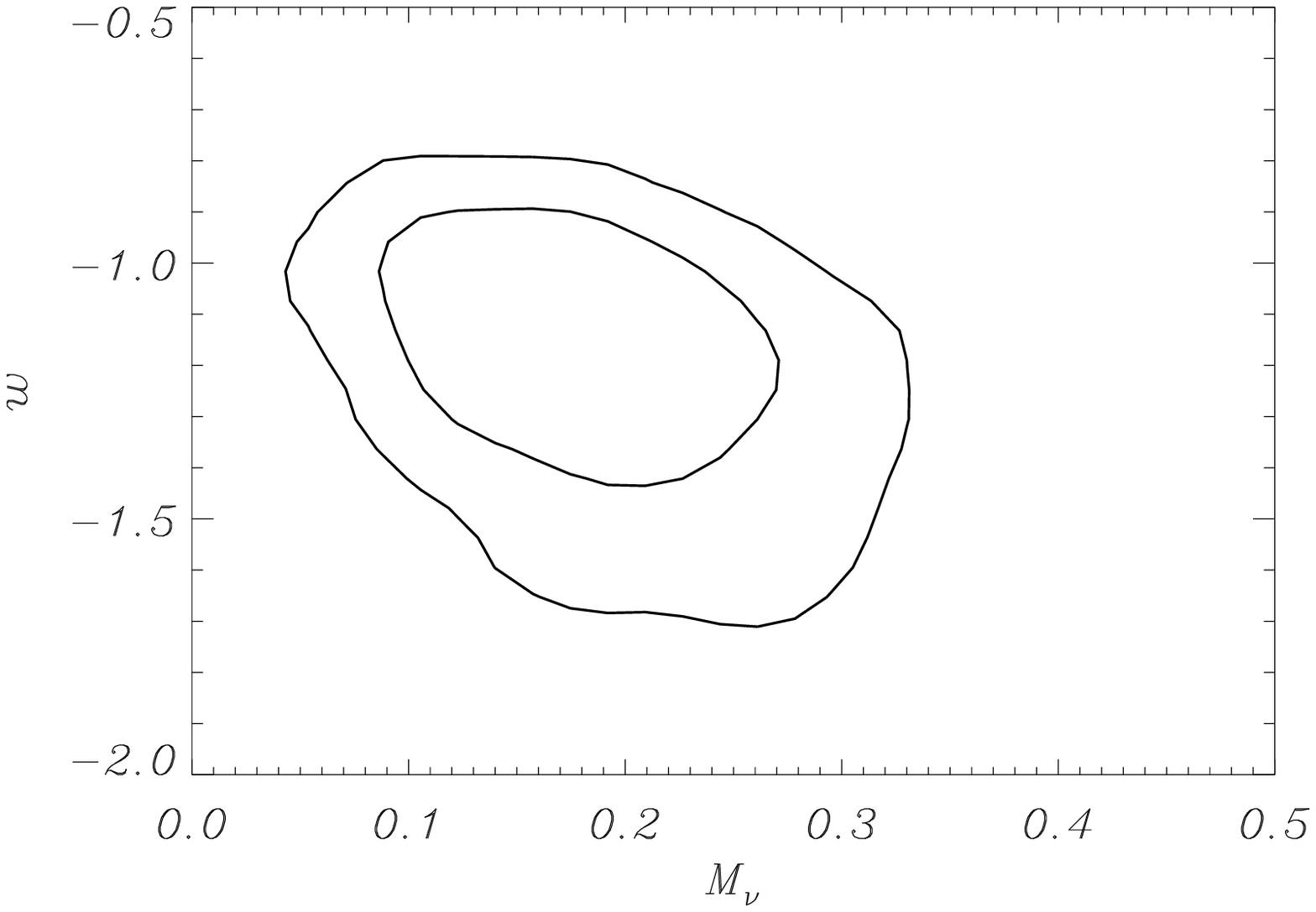}
\includegraphics[width=0.5\linewidth]{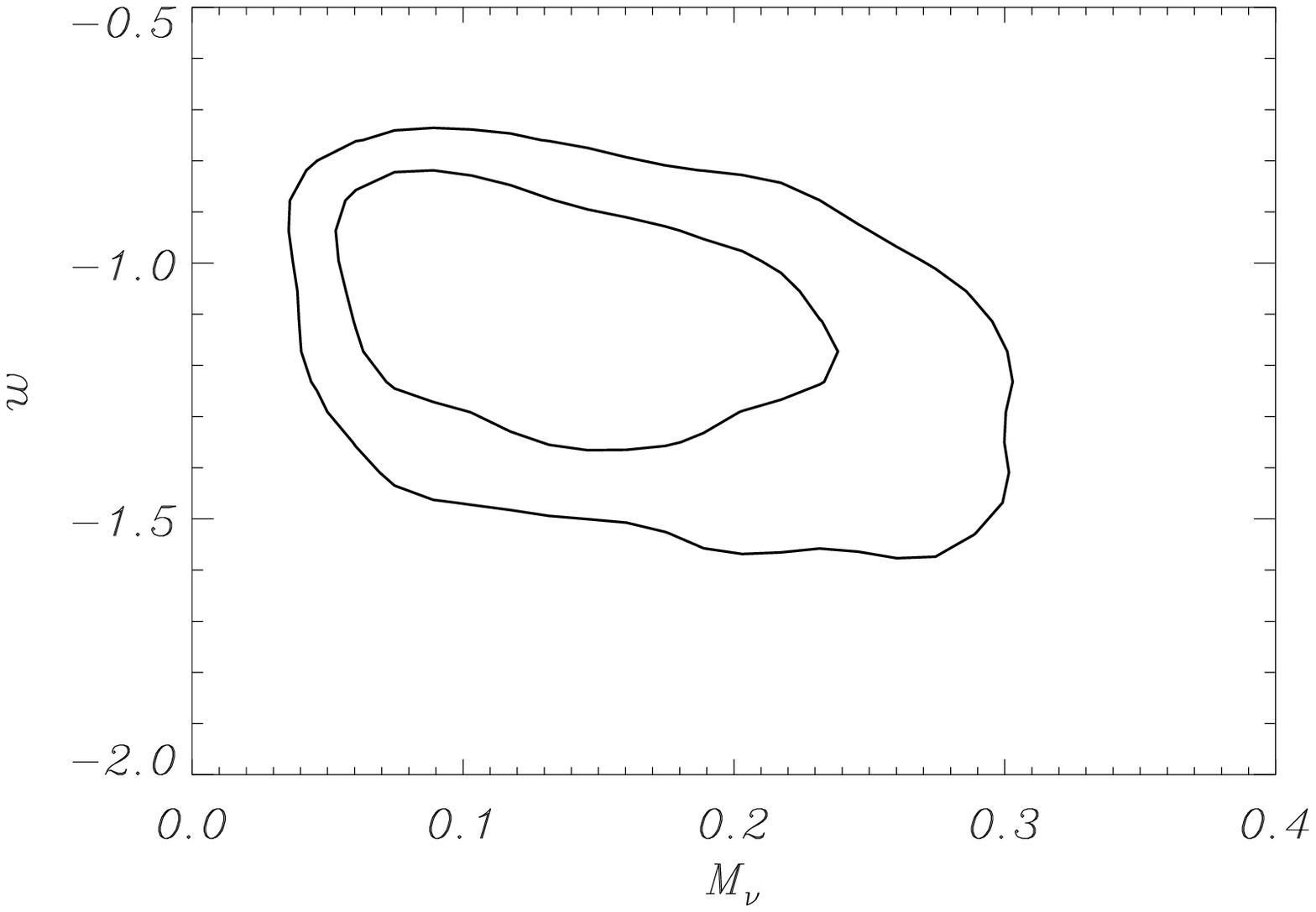}
\includegraphics[width=0.5\linewidth]{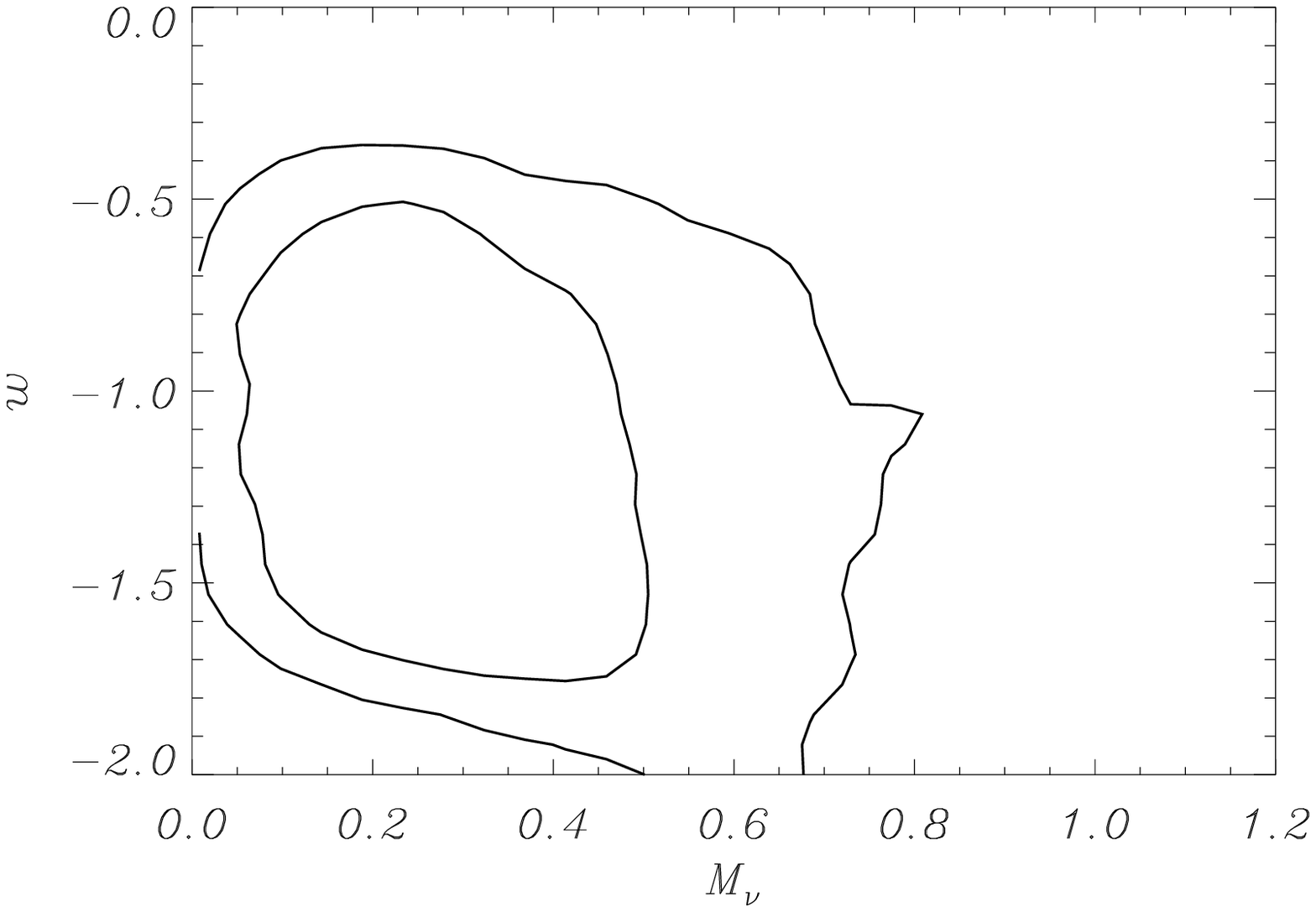}
\includegraphics[width=0.5\linewidth]{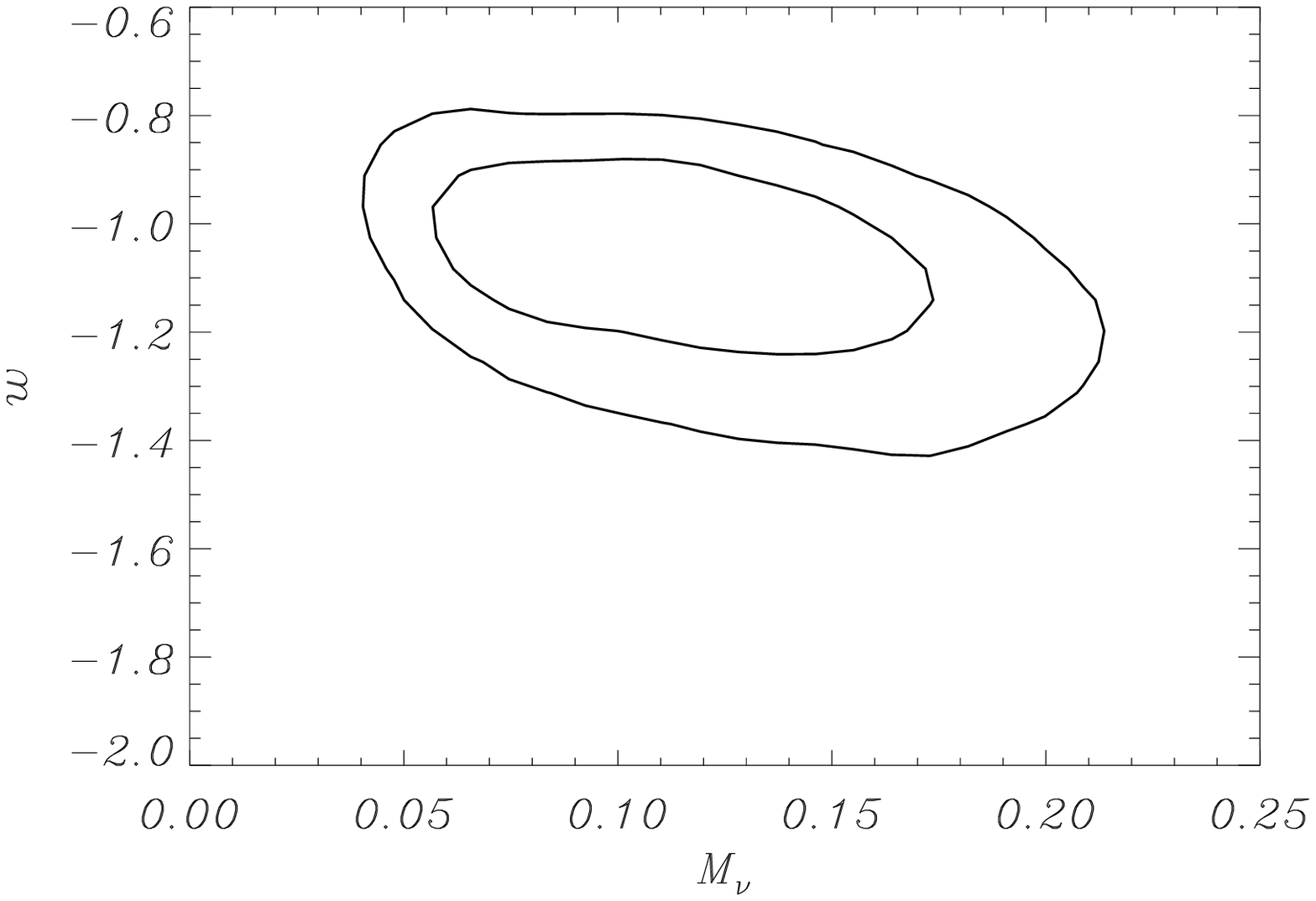}
\caption{The $M_{\nu}$-$w$ degeneracy: Shown are the results 
from PLANCK $\xi_\nu = 0$ (top left), $\xi_\nu \neq 0$ (top right), 
POLARBEAR $\xi_\nu \neq 0$ (bottom left) 
and EPIC $\xi_\nu \neq 0$ (bottom right) simulations. 
In this plot and each successive plot, the contours 
correspond to the 1- and 2-$\sigma$ regions. \label{fig:mnu-w}. 
The fiducial cosmological model is WMAP best-fit data and the 
neutrino masses $m_{2}$ and $m_{3}$ subject to neutrino 
oscillation results with $m_{1}$ assumed $0.01$eV.}
\end{figure*}

\begin{figure*}
\includegraphics[width=0.5\linewidth]{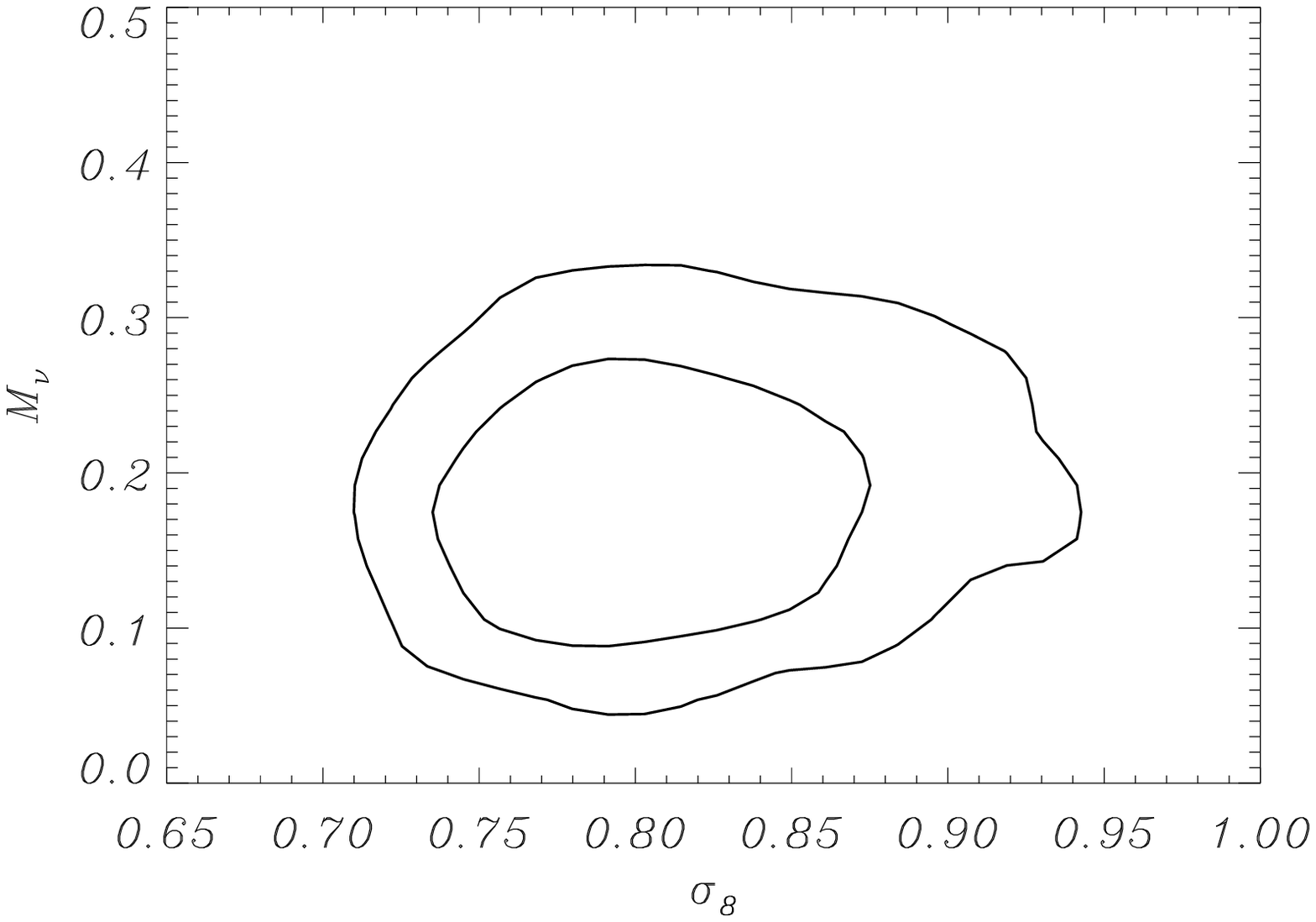}
\includegraphics[width=0.5\linewidth]{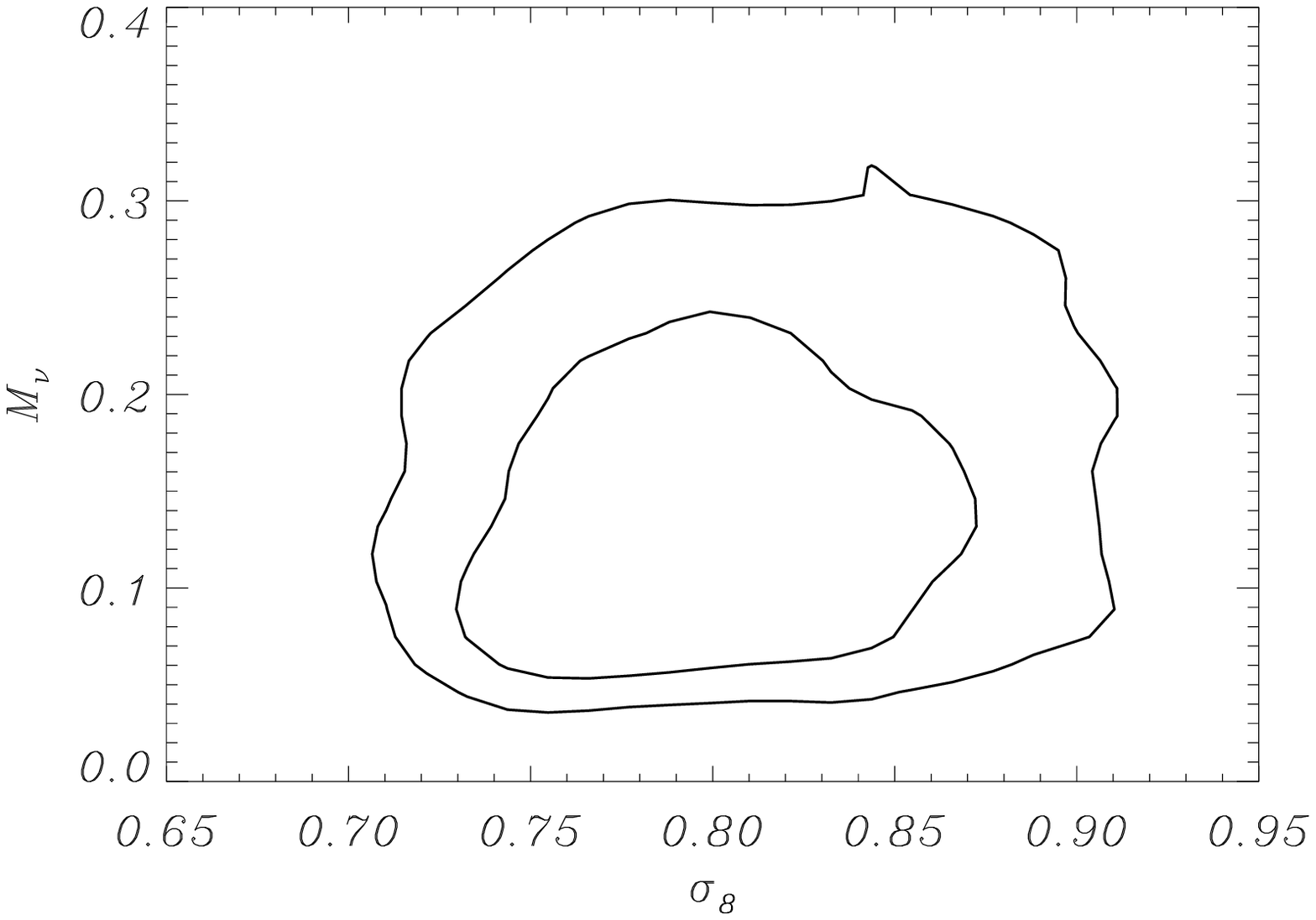}
\caption{The $M_{\nu}$-$\sigma_8$ degeneracy: 
Shown are the results for PLANCK in the case 
$\xi_{\nu}=0$ (left) and $\xi_{\nu}\neq 0$ (right). 
The fiducial cosmological model is as described in Fig. 5. 
\label{fig:mnu-sigma8}}
\end{figure*}

\begin{figure*}
\includegraphics[width=0.5\linewidth]{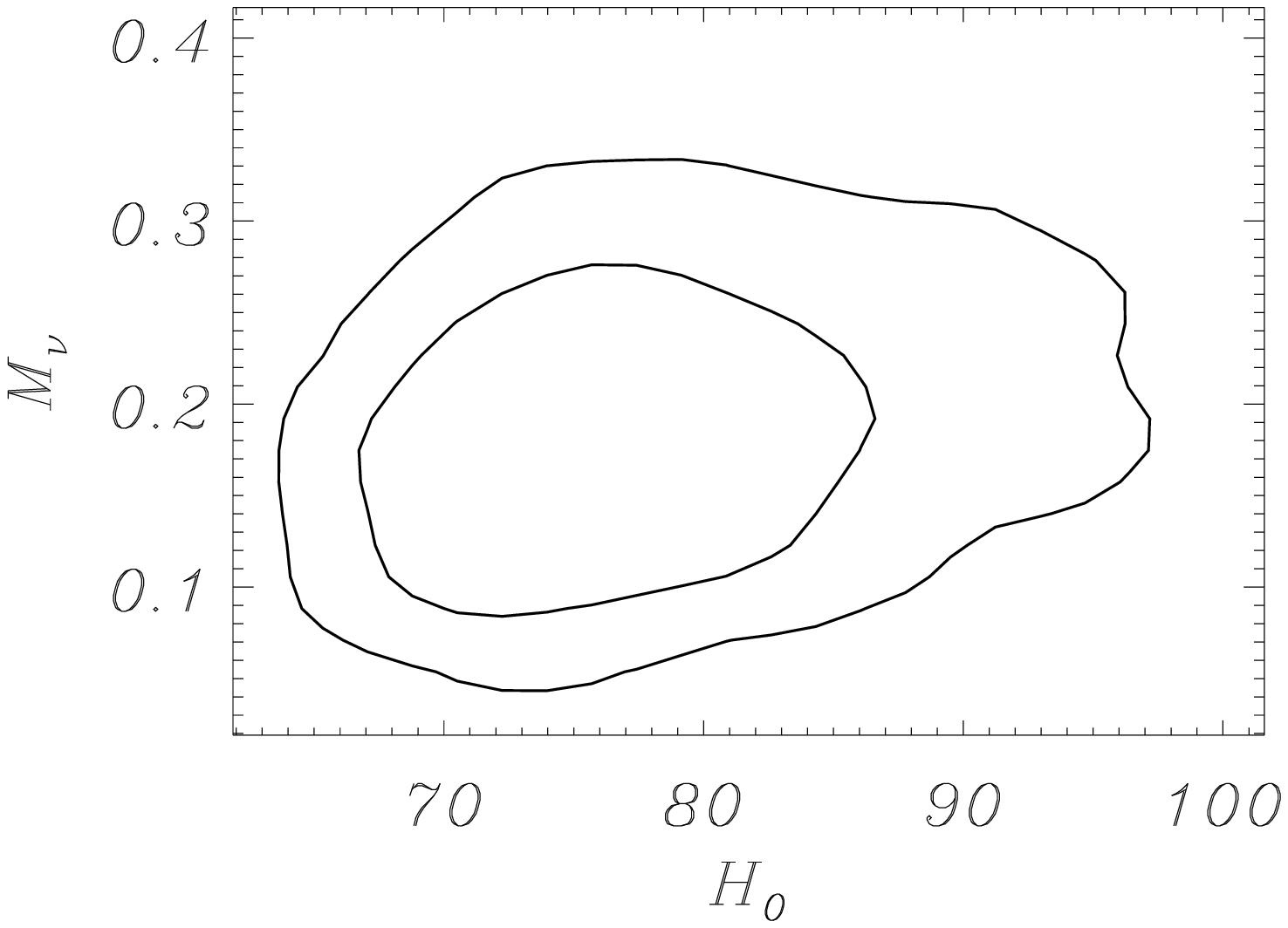}
\includegraphics[width=0.5\linewidth]{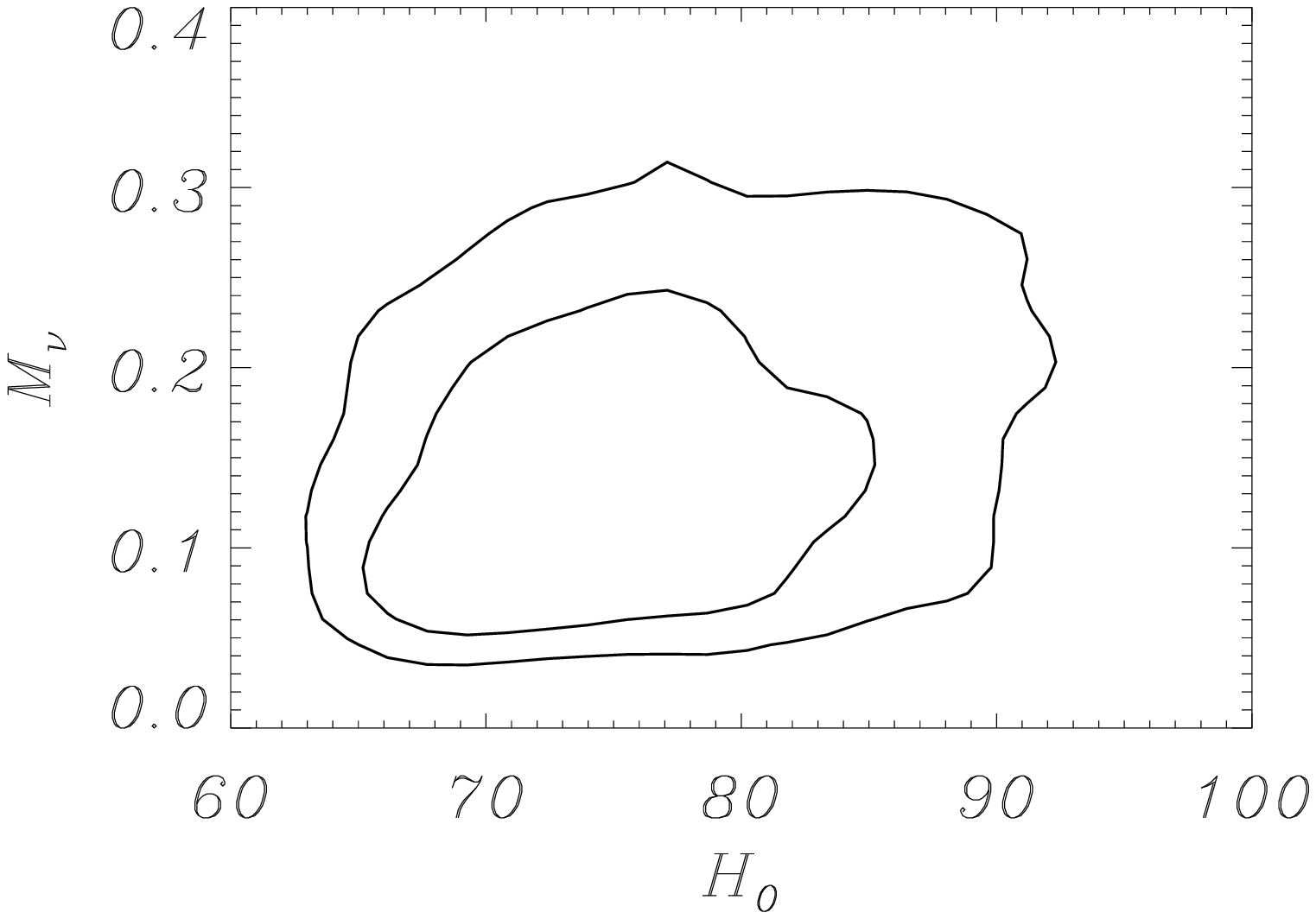}
\includegraphics[width=0.5\linewidth]{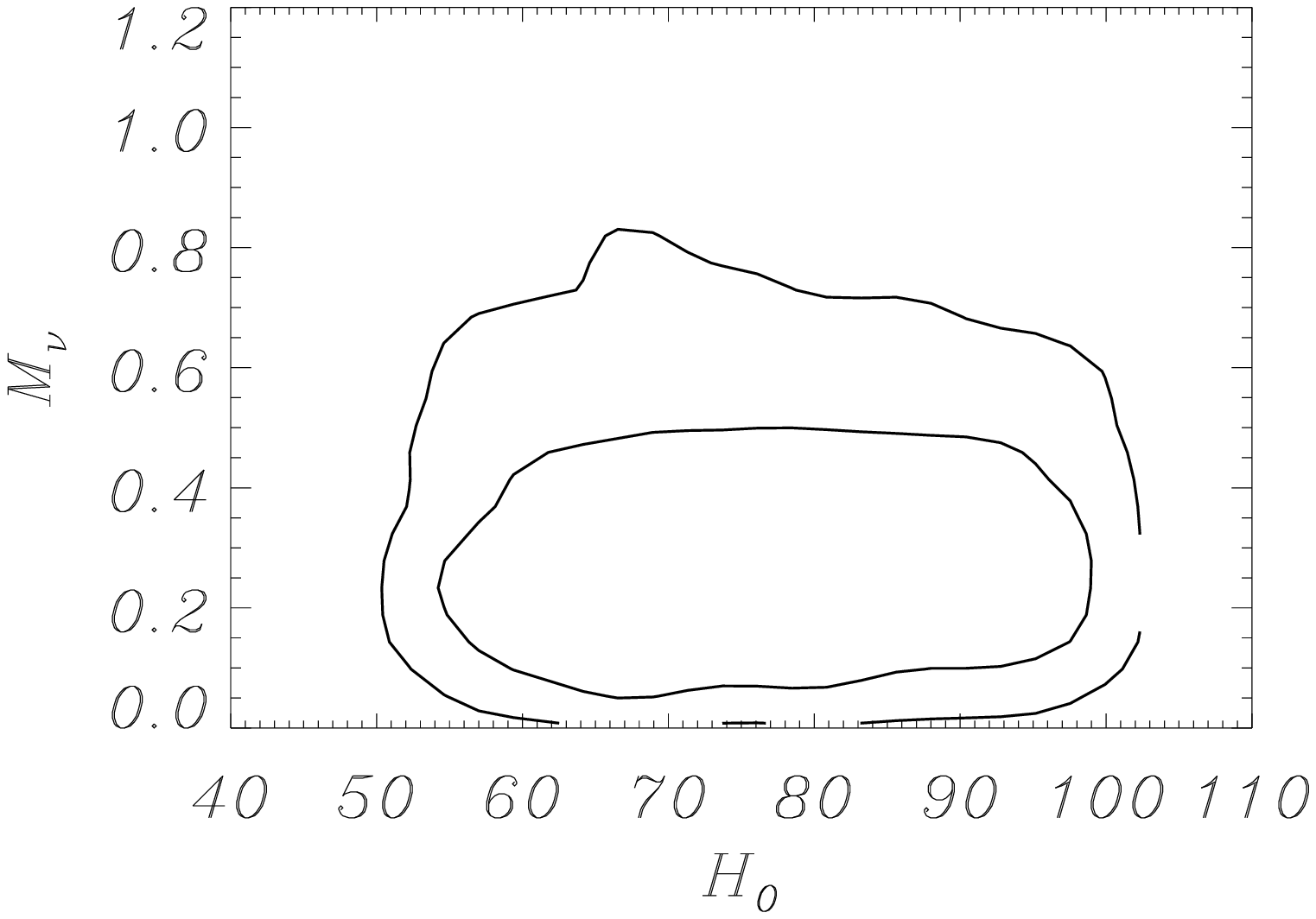}
\includegraphics[width=0.5\linewidth]{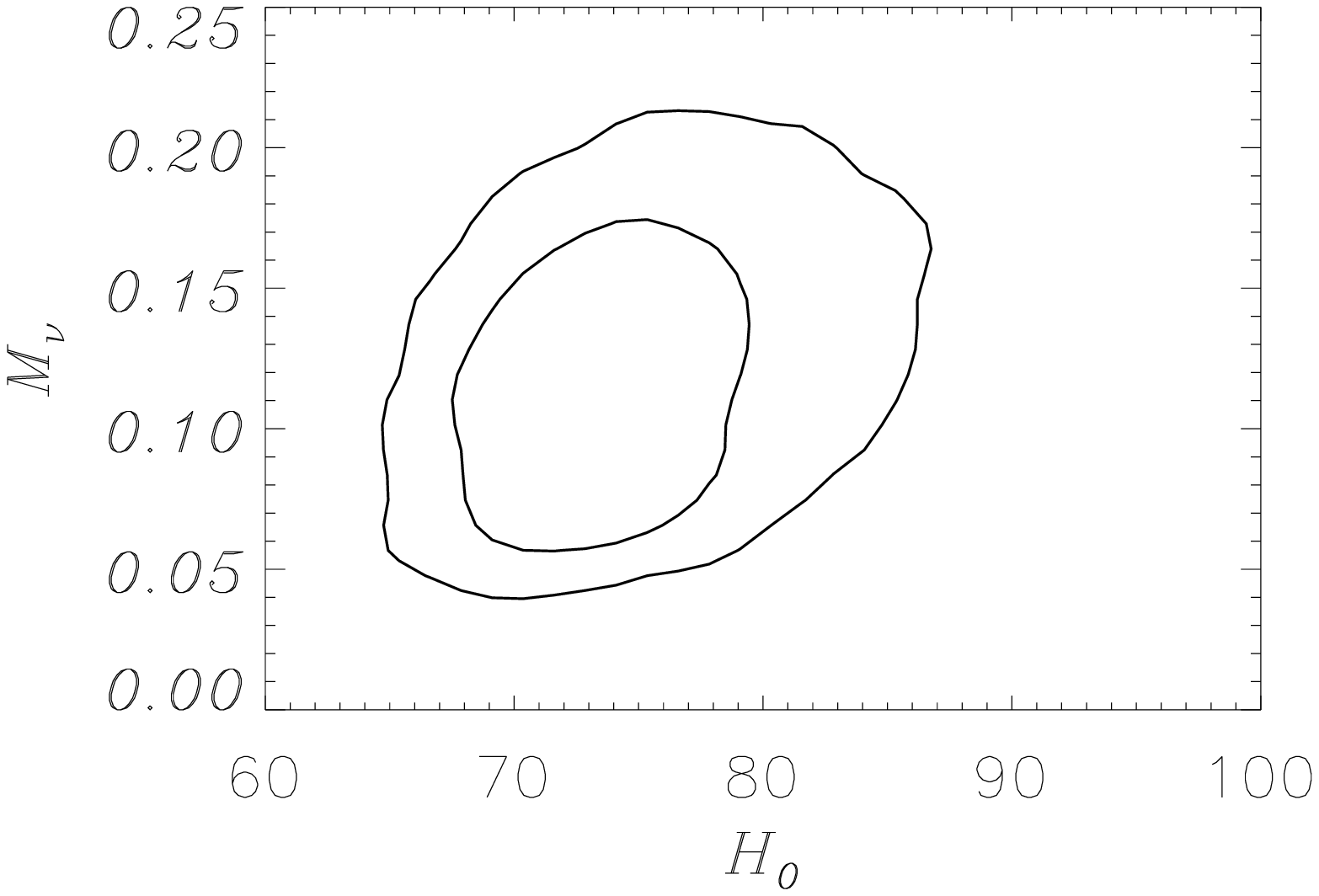}
\caption{The $M_{\nu}$-$H_{0}$ degeneracy: 
Shown are PLANCK ($\xi=0$ and $\xi\neq 0$ 
cases on the top-left and top-right, respectively), 
POLARBEAR (left bottom) and EPIC (right bottom) results. 
For POLARBEAR and EPIC we show the generalized cosmological 
model with $\xi_{\nu}\neq 0$. The fiducial cosmological 
model is as described in Fig. 5. \label{fig:mnu-h0}}
\end{figure*}

\begin{figure*}
\includegraphics[width=0.6\linewidth]{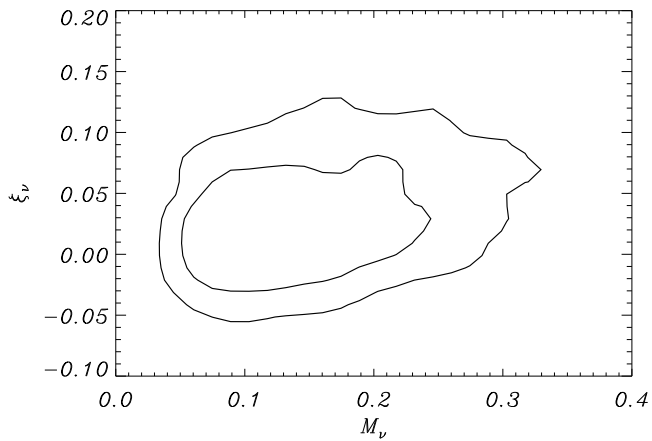}
\caption{The $M_{\nu}$-$\xi_{\nu}$ degeneracy for PLANCK: 
The fiducial cosmological model is as described in Fig. 5. 
\label{fig:mnu-xi}}
\end{figure*}

\begin{figure*}
\includegraphics[width=0.5\linewidth]{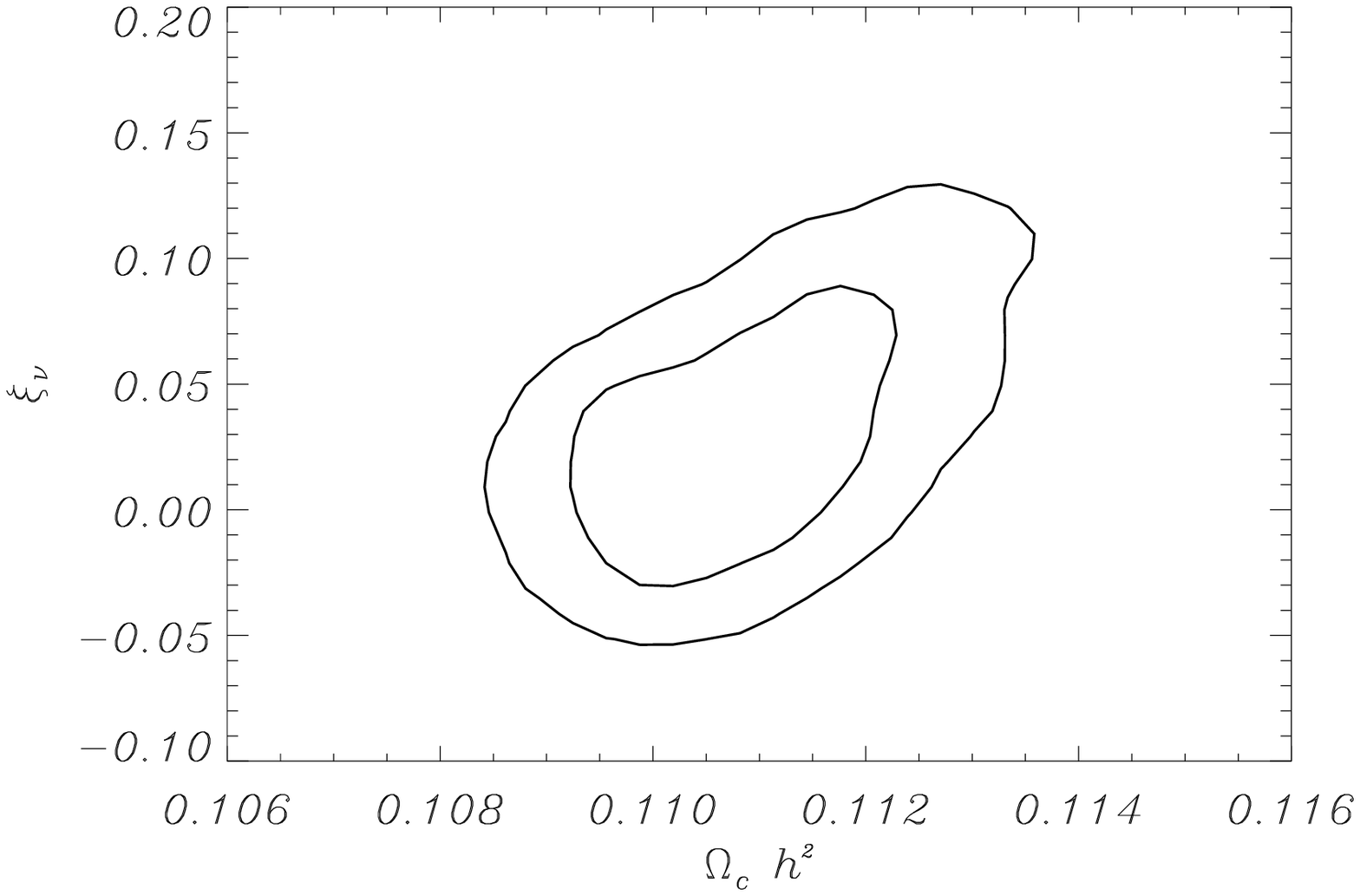}
\includegraphics[width=0.5\linewidth]{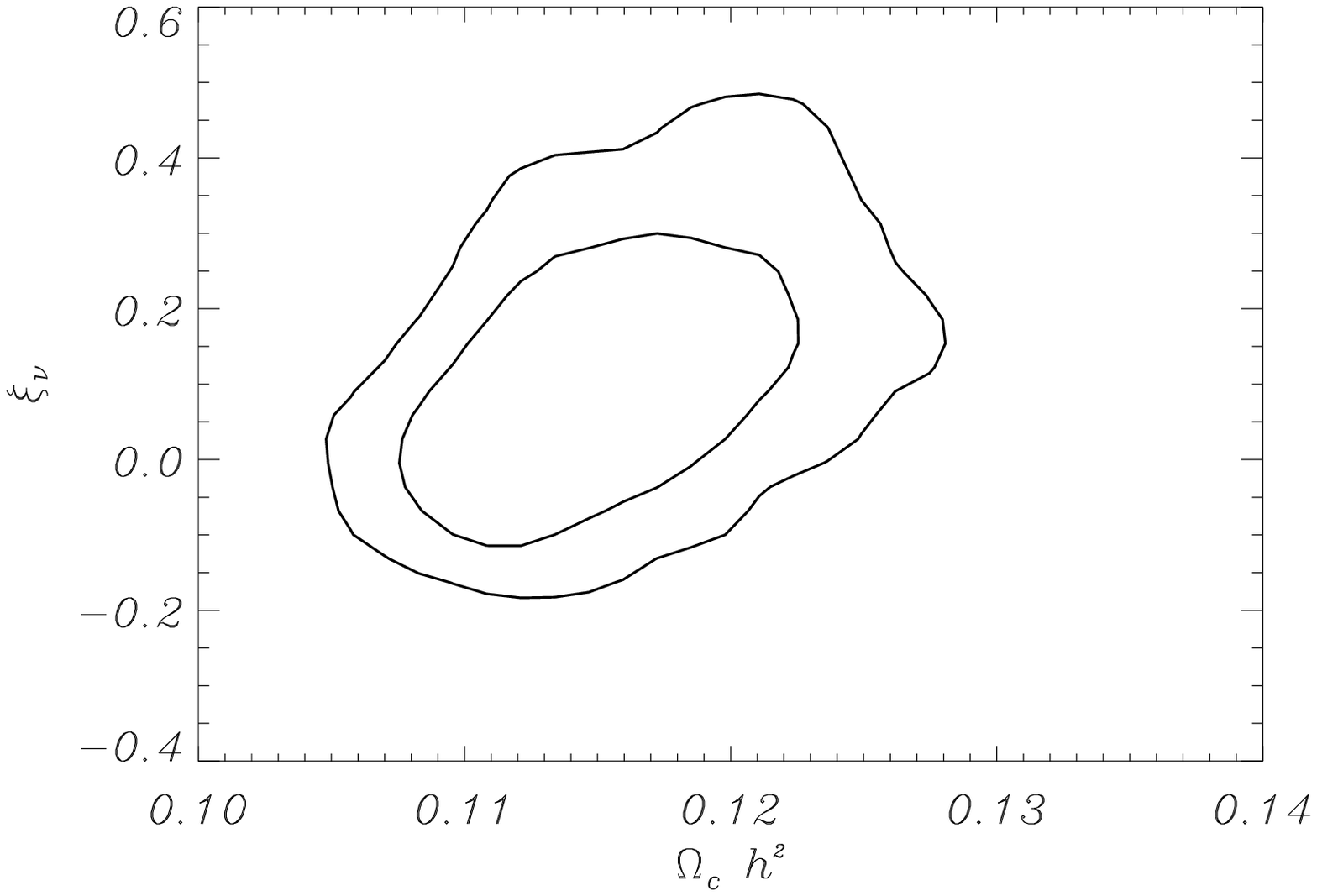}
\includegraphics[width=0.5\linewidth]{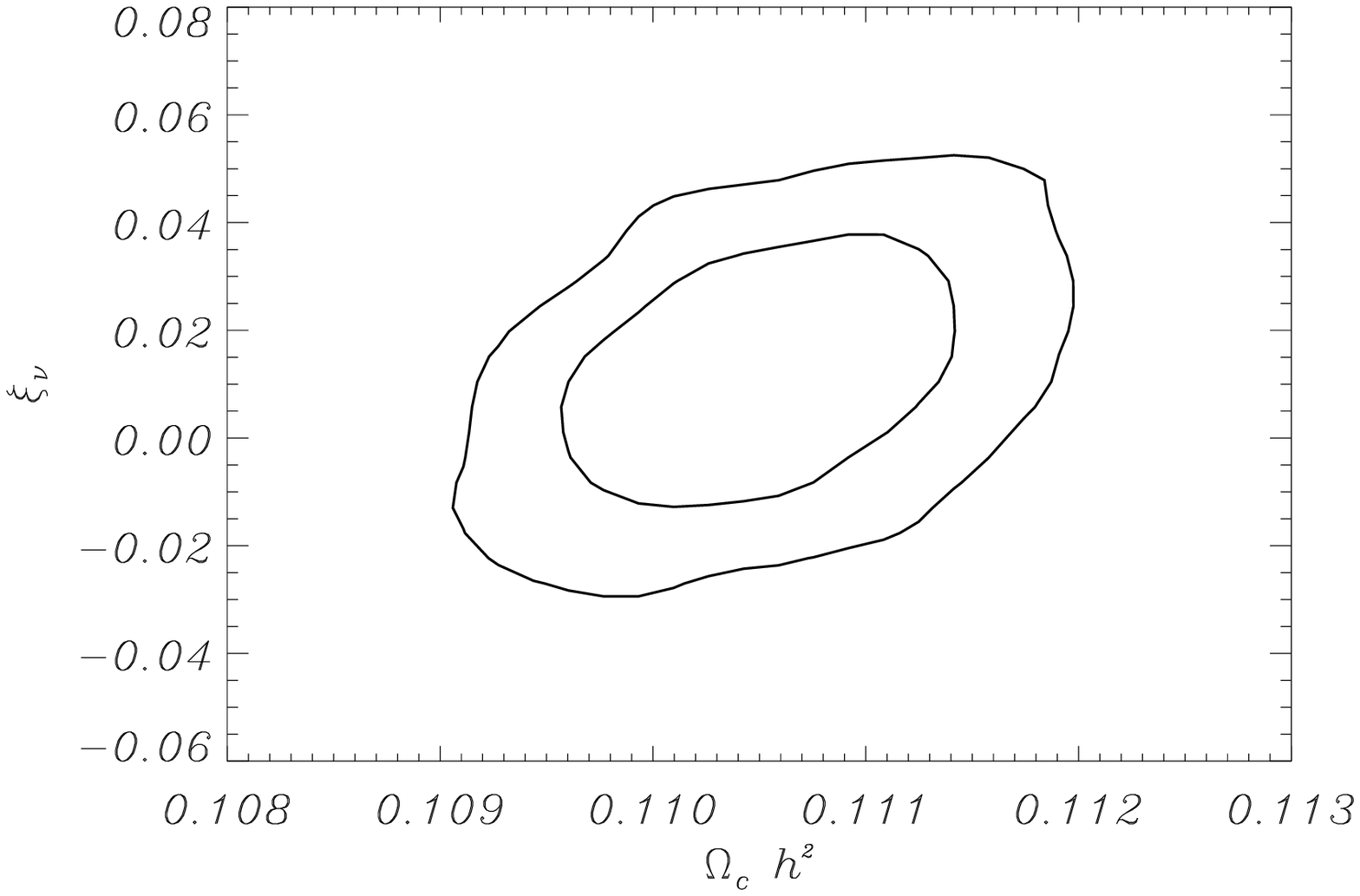}
\caption{The $\xi_{\nu}-\Omega_{c}h^{2}$ degeneracy: 
Shown are results for PLANCK (top left), POLARBEAR (top right) 
and EPIC (bottom). The fiducial cosmological model is as 
described in Fig. 5. \label{fig:xi-cdm}}
\end{figure*}

\begin{figure*}
\includegraphics[width=0.5\linewidth]{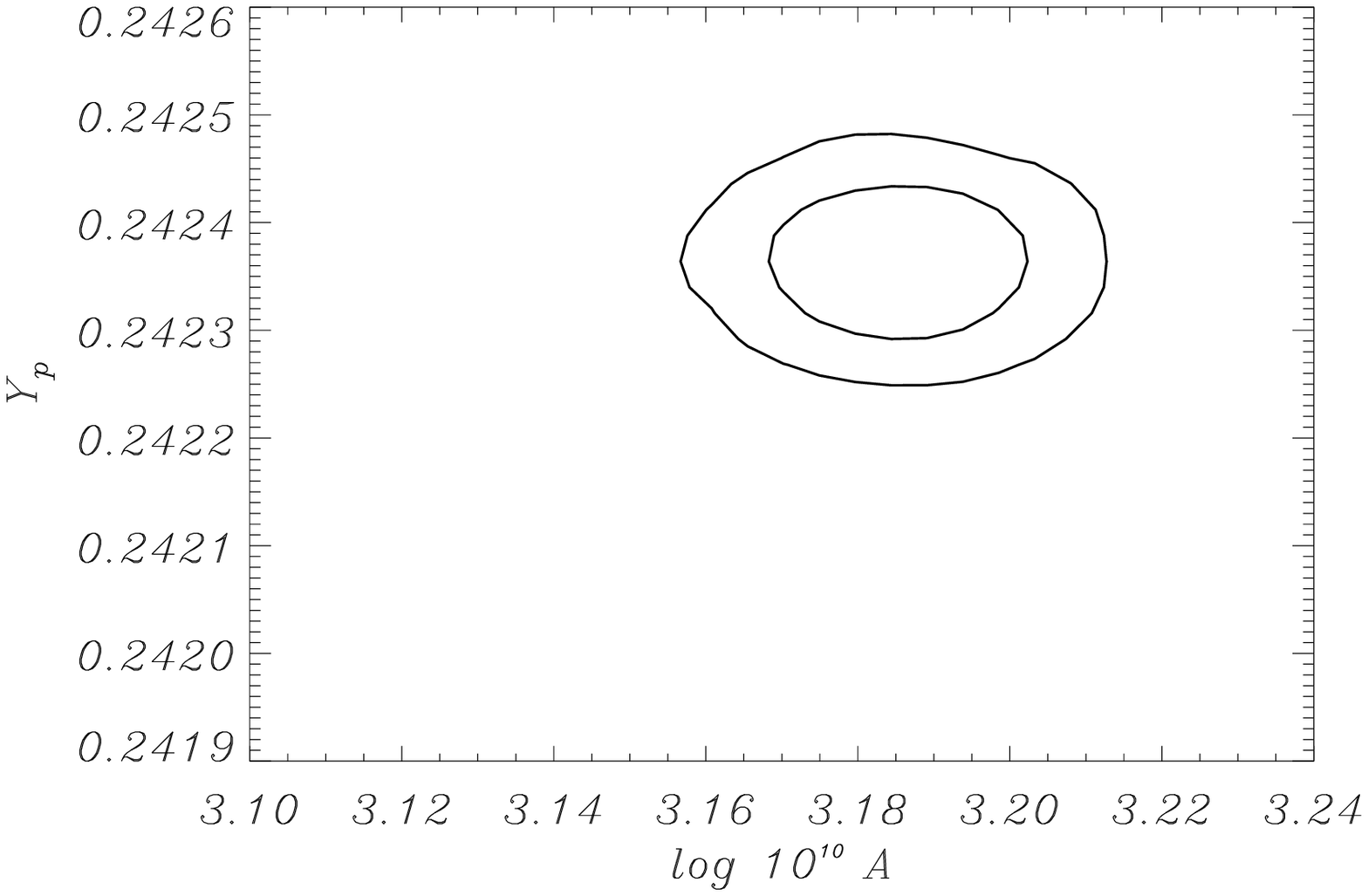}
\includegraphics[width=0.5\linewidth]{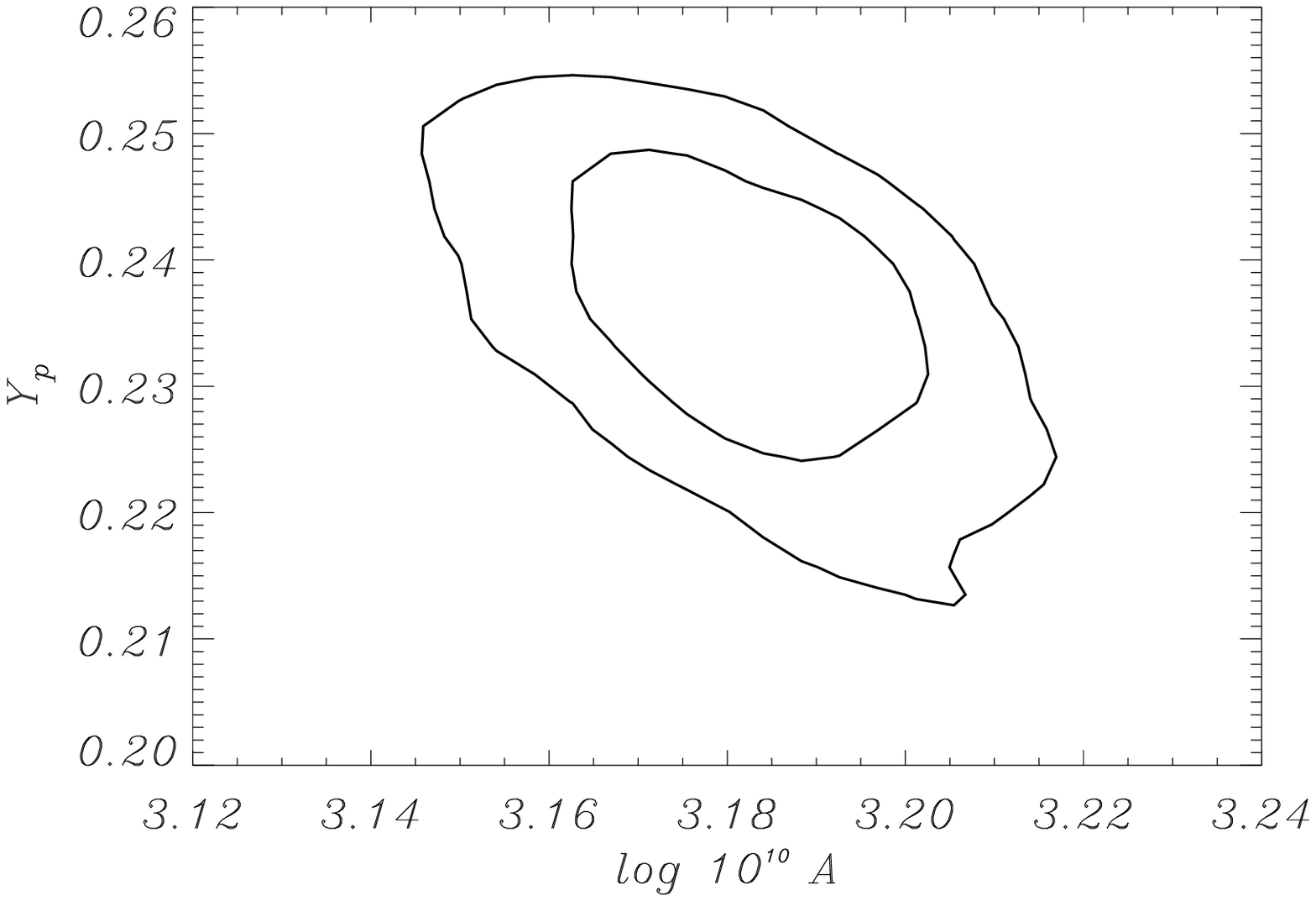}
\includegraphics[width=0.5\linewidth]{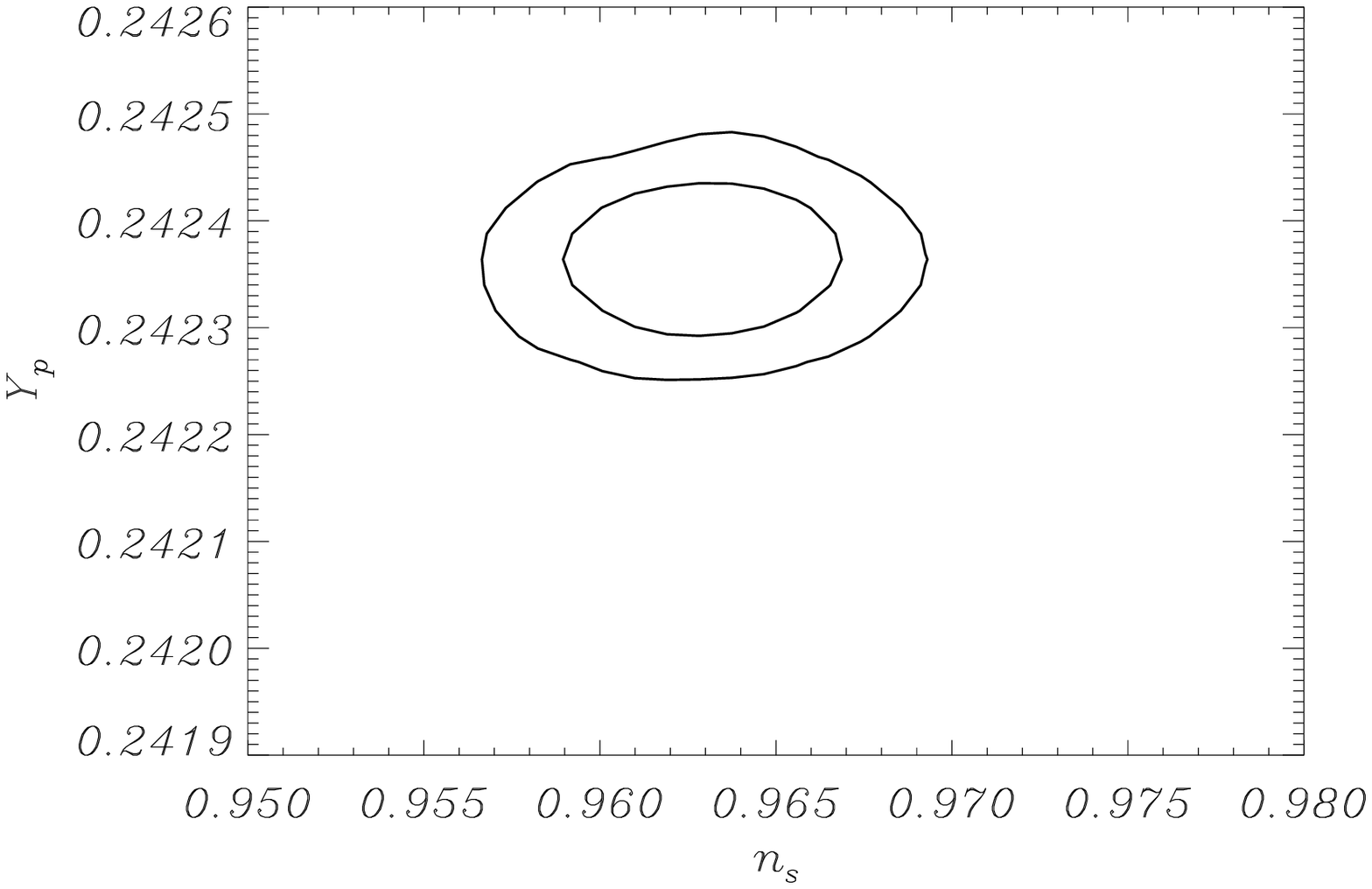}
\includegraphics[width=0.5\linewidth]{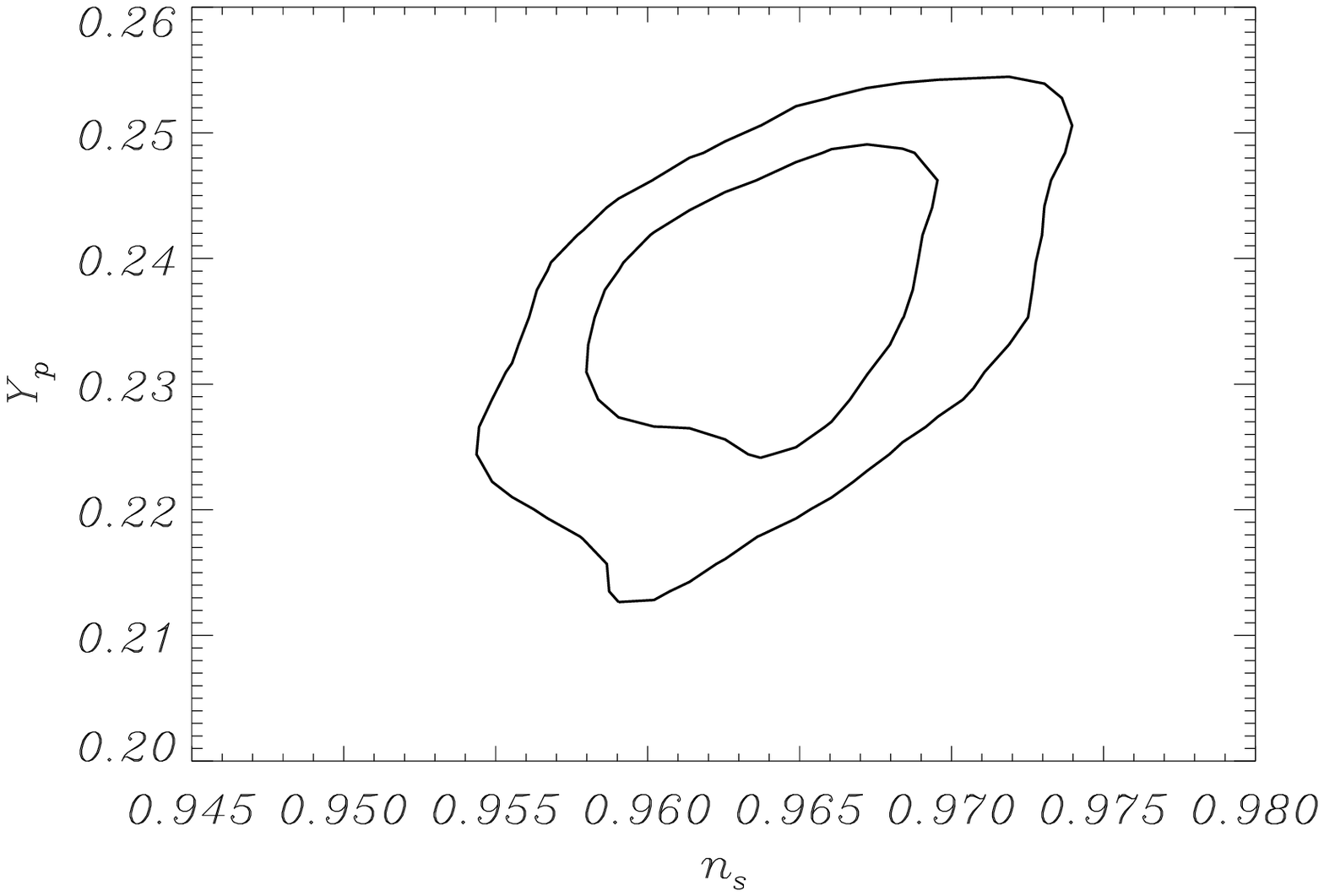}
\caption{The degeneracy of $Y_p$ with the normalization and 
tilt of the primordial power spectrum. The results for PLANCK 
with $\xi_\nu = 0$ are on the left and with $\xi_\nu \neq 0$ 
are on the right. The top two plots depict the $Y_p$-$A_{s}$ 
degeneracy and bottom plots show the $Y_p$-$n_s$ degeneracy. 
The fiducial cosmological model is as described in Fig. 5. 
\label{fig:yp-powerspectrum}}
\end{figure*}

\begin{figure*}
\includegraphics[width=0.5\linewidth]{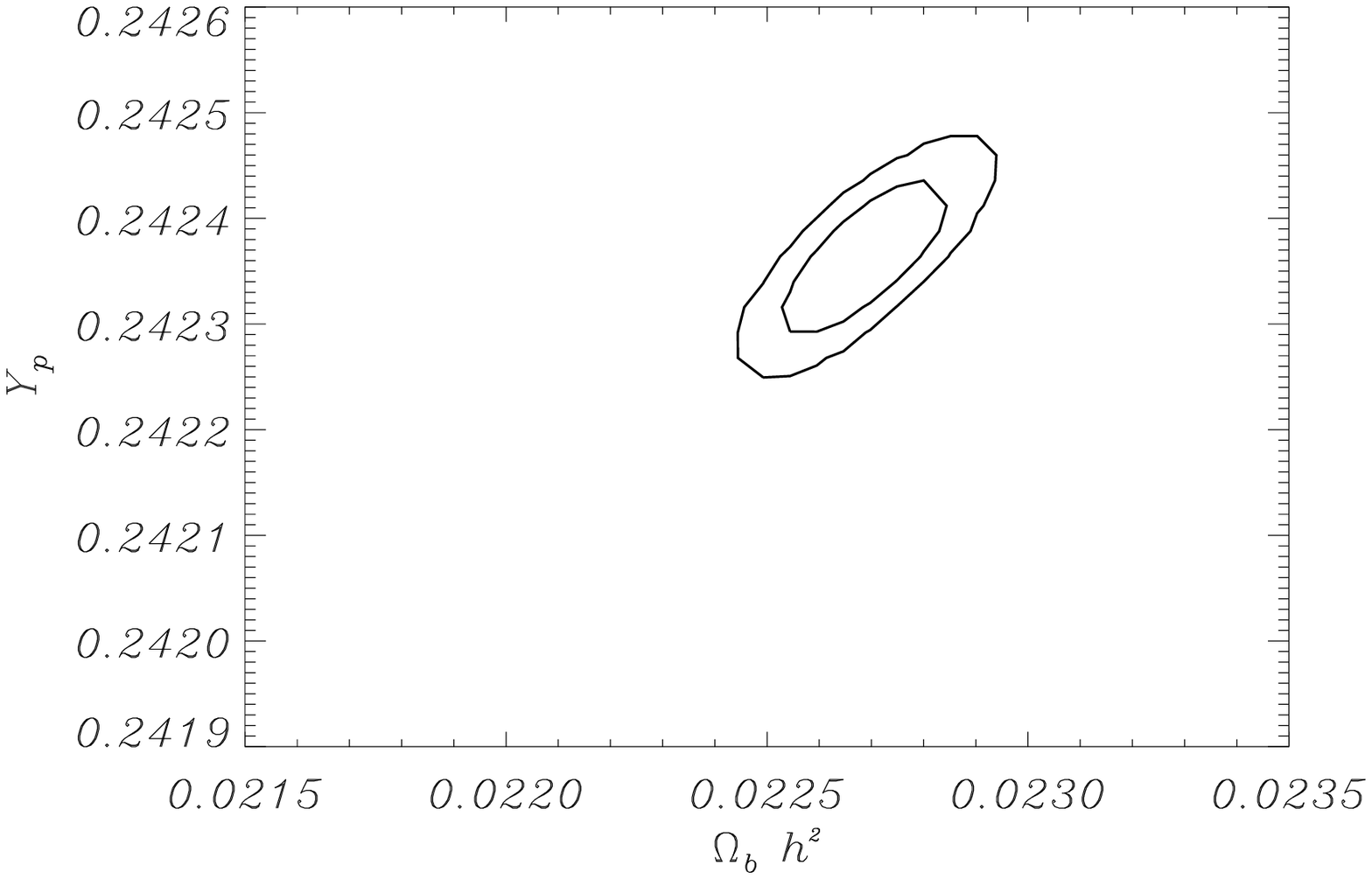}
\includegraphics[width=0.5\linewidth]{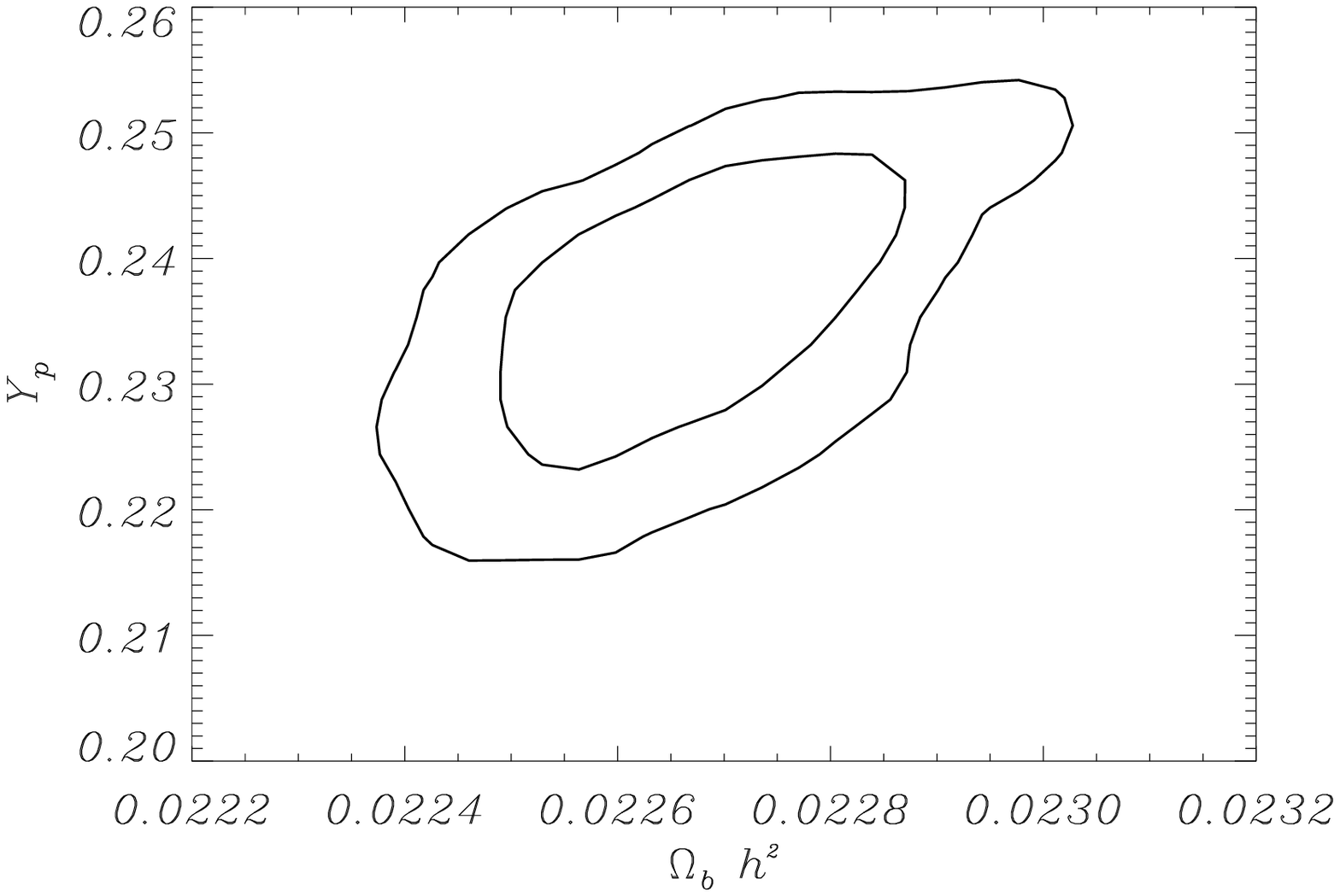}
\includegraphics[width=0.5\linewidth]{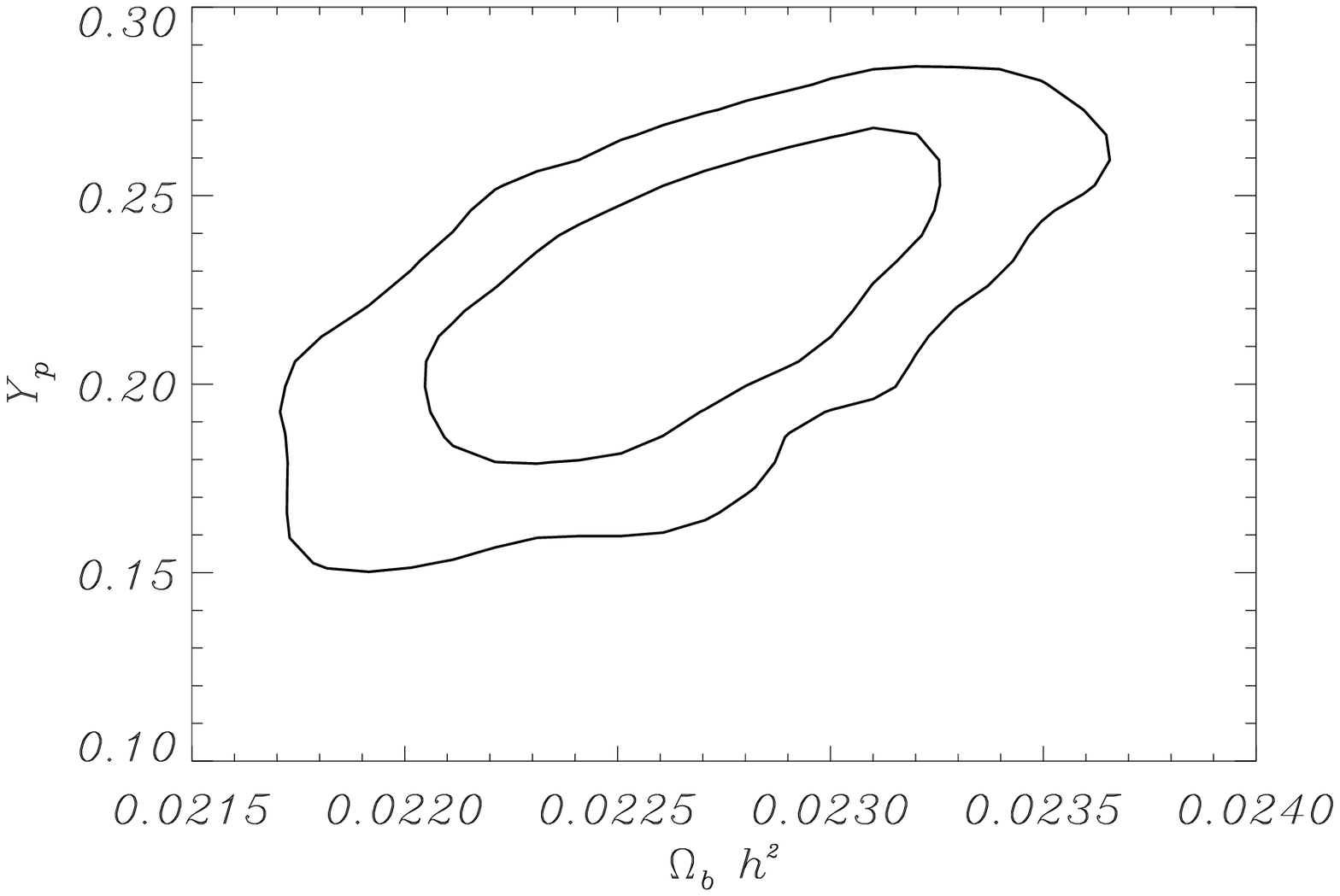}
\caption{The $Y_{p}-\Omega_{b}h^{2}$ degeneracy: Shown are results 
for PLANCK with $\xi_\nu = 0$  (top left), 
PLANCK with $\xi_\nu \neq 0$ (top right), and POLARBEAR with 
$\xi_\nu \neq 0$ (bottom). The fiducial cosmological model is 
as described in Fig. 5. \label{fig:yp-omegab}}
\end{figure*}

In Figures \ref{fig:mnu-w} - \ref{fig:yp-omegab}, we examine the degeneracies 
between different cosmological parameters and parameters that are affected by 
neutrino physics, in particular $M_\nu$, $\xi_\nu$ and $Y_p$.  In these figures, 
we assume the normal mass hierarchy with $m_1 = 0.01 ~{\rm eV}$, which corresponds 
to $M_\nu=0.073 ~{\rm eV}$ when the priors on the neutrino mass squared differences 
mentioned in the previous section are imposed, and $\xi_\nu = 0$ when calculating 
the theoretically expected CMB power spectra.  
All other cosmological parameters are set to the best fit values of 
WMAP \cite{komatsu.2009}. As a result, these figures illustrate the possible 
constraints that could be placed on the neutrino parameters if the actual values 
of these parameters are too small to create an effect on the CMB that could be 
discernible in the upcoming CMB experiments.  One important point to keep in mind 
is that the neutrino mass-squared differences measured in the laboratory excludes any 
$M_\nu \lesssim 0.058 ~{\rm eV}$.

\subsection{Neutrino Mass Degeneracy}

Cosmological probes of neutrino masses 
are sensitive to the kinematics of individual neutrinos. 
Since the gravitational interaction is flavor-blind, 
it does not distinguish between 
neutrino species.  However, for a fixed total mass 
it does depend on how this mass is distributed between 
the three species \cite{lesgourgues.1999}. 

\subsubsection{Degeneracy with $w$}

As mentioned above, the suppression of the matter power spectrum 
in the presence of massless neutrinos on scales much smaller than 
the neutrino free-streaming scale is\\
$\Delta P_{m}(k)/P_{m}(k)\approx -8\Omega_{\nu}/\Omega_{m}$.
Thus, increasing $\Omega_{\nu}$ as a result 
of increasing the neutrino mass can be compensated by 
increasing $\Omega_{m}$. However, to keep the universe 
spatially flat, the closure fraction of the dark energy 
must be lowered -- this is achieved by forcing the dark energy 
equation of state parameter, $w$, to be more negative. Therefore, 
increasing $m_{\nu}$ is degenerate with lowering 
$w$ as illustrated in Figure \ref{fig:mnu-w}. 
One way to avoid this degeneracy is to ignore cosmological 
information from scales smaller than the neutrino damping scale 
(which comes at the cost of significantly weakening the  
power of the CMB as a diagnostic tool of neutrino properties). 
Another possibility to avoid the $M_{\nu}-w$ degeneracy is to 
employ supplementary measures of distance, e.g., BAO or SNIa 
\cite{komatsu.2009}.

\subsubsection{Degeneracy with $\sigma_{8}$}

The fluctuation in the matter density on $8 h^{-1}~{\rm Mpc}$ scales is
\begin{eqnarray}
\sigma_{8}=\frac{1}{2\pi^{2}}\int P_{m}(k)W(kR)k^2dk ,
\end{eqnarray}
where $W(kR)$ is a window function, $h=H_0/(100 ~{\rm km}/{\rm s}/{\rm Mpc})$, 
and $R = 8 h^{-1} ~{\rm Mpc}$. Therefore, $\sigma_{8}$ is a function of both 
$A_{s}$ and $n_s$, the normalization and tilt of the power spectrum, 
respectively.  It is also a function of neutrino masses and degeneracy 
parameters, as well as any other cosmological parameters which may affect 
structure formation and the evolution of LSS on scales smaller than a few Mpc 
(Eq.\ 2.12).
Since $\sigma_{8}$ represents the mass fluctuation on $\sim 10 ~{\rm Mpc}$ 
scales, and is therefore subject to neutrino free-streaming we can expect a 
slight $M_{\nu}-\sigma_{8}$ degeneracy, at least for CMB experiments which are 
sensitive to angular scales that correspond to neutrino 
free streaming scales. However, this degeneracy is very weak in practice 
as can be seen from Figure 6; no such degeneracy is expected to be observed 
in the PLANCK data.

\subsubsection{Degeneracy with $H_{0}$}

Previous studies have shown an anti-correlation between neutrino mass and 
the Hubble constant. The anti-correlation results from the fact that while 
all three neutrino mass states are at least mildly-relativistic at 
recombination (consistent 
with the WMAP constraints on neutrino mass \cite{komatsu.2009}), at least 
two of these mass states are non-relativistic today (consistent with the 
$\delta m^2$ values from neutrino experiments).  As a consequence, the 
neutrinos contribute to $\Omega_m$ today, but contributed to $\Omega_r$ 
(the closure fraction in radiation energy density)
at recombination.  Since neutrinos with larger rest masses will constitute 
a larger fraction of the CDM at the current epoch, larger 
neutrino masses imply a larger $\Omega_r$, relative to $\Omega_m$, on the 
surface of last scattering which, in turn, gives an enhanced ISW effect.
Most of the extra power due to this effect is on scales somewhat larger 
than the horizon (the first acoustic peak), effectively extending the 
first acoustic peak to larger scales.  This effect can be mimicked by 
lowering $H_0$, because a lower $H_0$ implies a larger horizon at decoupling.

Figure \ref{fig:mnu-h0} shows the degeneracies between $M_\nu$ and $H_0$ for 
PLANCK, POLARBEAR, and EPIC.  In these plots, the $M_\nu - H_0$ degeneracy 
described above is not evident.  This results from the fact that these high 
resolution experiments will constrain neutrino masses primarily from lensing 
information, instead of through the ISW effect at low multipoles.  For these 
high resolution experiments, the neutrino free streaming length is the 
more relevant quantity that relates to observables. This is an example 
of how the higher resolution and sensitivities of upcoming CMB experiments 
open windows to new effects that may lift parameter degeneracies.

\subsection{Neutrino Chemical Potential Degeneracy} 

Including nonzero neutrino degeneracy parameters in the analysis introduces new 
parameter degeneracies.  In Figure \ref{fig:mnu-xi}, the degeneracy between 
$\xi_\nu$ and $M_\nu$ is shown.  A naive interpretation is 
that for a given neutrino free streaming length, increasing $\xi_\nu$ must 
be compensated by increasing the neutrino mass.  However, the range of allowed 
neutrino masses and chemical potentials does not allow such a parameter 
degeneracy in the neutrino free streaming scale 
(see also Fig.\ \ref{fig:freestreaming}). The mild degeneracy shown 
here comes from the physics at recombination through the decay of potential 
wells and the ISW effect. Nonvanishing neutrino degeneracy parameters increase 
the energy density in neutrinos for fixed neutrino masses. 
Increasing the neutrino energy density must be compensated by increasing the 
density of CDM in order to keep 
$\Delta P_m / P_m \approx - 8 \Omega_\nu / \Omega_m$ unchanged.  
This degeneracy can be seen in Figure \ref{fig:xi-cdm}.

\subsection{Helium Fraction Degeneracy} 

The helium fraction affects the physics of recombination primarily
by changing the Silk damping scale.  
The baryon closure fraction, $\Omega_{b}$, is obtained to high precision 
from the amplitudes of the acoustic peaks of the CMB.
For a given $\Omega_{b}$, more helium implies 
less hydrogen and fewer free electrons on the surface of last scattering. 
This causes a larger photon mean free path, which damps CMB temperature anisotropy
on larger angular scales.  
This effect can be mimicked by either reducing the
normalization of the primordial power spectrum or increasing its tilt. 
Both the $Y_{p}-A_{s}$ and the $Y_{p}-n_{s}$ planes 
are shown in Figure \ref{fig:yp-powerspectrum}.  There is a significant difference 
in the $Y_p$ axes between the $\xi_{\nu}=0$ and the $\xi_{\nu}\neq 0$ cases; BBN 
data and the current precision on cosmological parameters tightly constrain $Y_p$, 
but allowing $\xi_\nu$ to be nonzero affects the BBN-calculated yield for $Y_p$ 
and allows these degeneracies to manifest themselves in the analysis. The dilution 
of the free electron density at the epoch of last scattering also can be
compensated by increasing $\Omega_{b}h^{2}$ as can be seen in Figure \ref{fig:yp-omegab}.

\section{Results}

We adopt a 14 parameter cosmological model with priors on neutrino masses taken 
from neutrino oscillation data.  Our model is consistent with the concordance 
cosmological model \cite{komatsu.2009}.  
With this model, a BBN+CMB MCMC analysis demonstrates that
PLANCK and POLARBEAR will be able to measure a total neutrino mass of $0.29 ~{\rm eV}$ 
(PLANCK) and $0.75 ~{\rm eV}$ (POLARBEAR) at the 95\% confidence level.
In addition, neutrino degeneracy parameters can be constrained 
to be smaller than $0.11$ ($\xi_e$) and $0.49$ ($\xi_{\mu}$, $\xi_{\tau}$) for 
PLANCK and $0.62$ ($\xi_e$) and $1.1$ ($\xi_{\mu}$, $\xi_{\tau}$) for POLARBEAR.
The former constraint on $\xi_{\nu_{e}}$ is already better than the corresponding 
BBN one, Eq.\ (2.7).

It is interesting to examine the sensitivities of upcoming CMB experiments to 
the neutral lepton asymmetries. If the 
neutrino asymmetries equilibrated through neutrino oscillations prior to the BBN epoch, 
then the possible constraints on the neutrino degeneracy parameters become very strong, 
$\xi_{\nu_e} = \xi_{\nu_\mu} = \xi_{\nu_\tau} < 0.06$ \cite{Serpico.2005}.

Our analysis with the extended parameter space yields 
weaker constraints on neutrino masses. For example, 
in the minimal model (all $\xi_{\nu}$ are set to $0$) the PLANCK
$2\sigma$ upper limit on $M_\nu$ is $0.27 ~{\rm eV}$. When the degeneracy 
parameters are turned on, the corresponding upper 
limit on $M_{\nu}$ rises to $0.29 ~{\rm eV}$.
The $1\sigma$ confidence range on the electron neutrino degeneracy parameters are
$-0.0314<\xi_{\nu_{e}}<0.108$ for PLANCK and 
$-0.017<\xi_{\nu_{e}}<0.33$ for POLARBEAR. The reason 
for the skewness of the distribution towards positive 
$\xi_{\nu_{e}}$ values results from the fact 
that $\nu_{e}$ determines the reaction rates of the processes described in Eq. (2.6) and also 
the expansion rate.
In contradistinction, $\xi_{\nu_{\mu}}$ 
and $\xi_{\nu_{\tau}}$ affect only the expansion rate.

Perhaps the ultimate CMB experiment to address B-mode related issues is the mission concept, 
EPIC. We find that EPIC will be able to set an upper limit on the total neutrino mass of 
$\sim 0.20~{\rm eV}$ at $2\sigma$ confidence.  In addition, we find that EPIC data alone 
will have sufficient sensitivity to constrain the neutrino degeneracy parameters to a level 
which can compete with the current constraints on degeneracy parameters derived from the 
primordial abundance of light elements.  We derive the following limits on the degeneracy parameters: 
$\xi_{\nu_e} < 0.045$ and $\xi_{\nu_{\mu,\tau}} < 0.29$ at $2\sigma$. These are better than 
current BBN constraints, even if equilibration of the degeneracy parameters is assumed.
These results are achievable without resorting to assumptions about flavor mixing in the 
early universe.  EPIC is capable of such an improvement in sensitivity to the degeneracy 
parameters because of its very high sensitivity and angular resolution which allow for 
the precise measurement of the B-mode polarization required for lensing extraction of the CMB.

Finally, a cautionary note on the efficacy of our analysis. 
It is reasonable to ask whether future CMB data will indeed warrant considering 
a 13- or 14-parameter model. This question is reasonable even in the context of idealized 
analysis presented here because CMB data is at least limited by cosmic variance and instrumental 
noise. In the real world it will also be limited by astrophysical foregrounds as well as systematics. 
As in \cite{Liddle.2004}, adding cosmological parameters is in general expected to improve the fit of data to 
the theoretical model. Defining a generalized $\chi^{2}$ such as 
$\tilde{\chi}^{2}=-2\ln(L)+2p$ (where $L$ is a likelihood 
and $p$ is the number of cosmological parameters in a given model) and exploring if it improves is 
a useful test for such models \cite{Liddle.2004}. However, this requires real data, i.e. sky-maps in the 
CMB case. Our analysis employed a mock power spectrum which in principle we could use to generate 
multiple sky realizations, each one yielding a different numerical value of $\tilde{\chi}^{2}$. 
One can then statistically determine what fraction of these actually improve when we extend the model 
from 11 to 13 or 14 parameters. However, this procedure is time consuming and may not be necessary at 
this point. When the real CMB data considered in this work is available it will be straightforward 
to determine whether or not our generalized model gives a better fit to the data.

\section{Conclusion}

Within a decade the CMB has transformed from being a 
cosmological probe of the basic cosmological parameters 
to a probe of physics beyond the standard model. 
CMB experiments have set interesting 
limits on the energy scale of inflation as well as on exotic physics 
such as topological defects from phase transitions in the 
early universe and cosmological birefringence.
Also, important
constraints on neutrino masses have already been obtained. 
Although WMAP has constrained neutrino masses to the sub-eV level, 
it is an exciting possibility that precise CMB measurements could place 
stringent constraints on neutrino masses and neutral lepton asymmetries.
Although the C$\nu$B neutrinos cannot be directly detected, they can be 
indirectly detected through their dynamics (through their effect on the 
expansion rate) and kinematics (via the damping of LSS by neutrino free 
streaming). A convincing detection of the C$\nu$B would be a monumental 
discovery in the history of cosmology.

In this paper we explored the effects of neutrino mass and nonzero neutrino 
degeneracy parameters on the CMB.  Changing $\xi_\nu$ leads to a different 
value of $Y_p$ from BBN, and $Y_p$ affects the density of free electrons at 
recombination. $Y_p$ was calculated self-consistently in our 
analysis by a BBN code with a given set of cosmological and neutrino 
parameters.  No priors on $Y_p$ were included or used in the analysis.

Our analysis is conservative in that it allows the three neutrino degeneracy 
parameters to be independent parameters in the analysis.  Typically, 
the neutrino degeneracy parameters are assumed to be equal to each other, 
which would be expected if the neutrino asymmetries equilibrated in 
the early universe.  Other works have at least set $\xi_{\nu_\mu} = \xi_{\nu_\tau}$, 
since the physics of the early universe is insensitive to the difference 
between these parameters.  However, neutrino free streaming lengths are 
sensitive to the absolute value of each neutrino degeneracy parameter, so 
we treated each degeneracy parameter independently.  While the addition of 
data from other cosmological probes of distance scales or LSS could be 
included to break parameter degeneracies, we did not include them so that 
we could isolate the probative powers of the CMB alone.

Upcoming CMB experiments such as PLANCK, POLARBEAR, and perhaps also EPIC will have 
the capability to either detect the neutrino masses and degeneracy 
parameters or place much more stringent bounds on these parameters as 
compared to current constraints from WMAP. These breakthroughs in the 
power of the CMB to detect neutrino parameters is the direct result of 
the improved resolution and sensitivity of upcoming CMB experiments, 
allowing CMB lensing extraction to provide an ultra-sensitive handle 
on neutrino masses and degeneracy parameters.

\section*{Acknowledgments}

We thank the referee for very helpful comments.
Eric Linder is acknowledged for his useful suggestions.
We acknowledge the use of the publically available code by Lesgourgues, 
Perotto, Pastor \& Piat for the calculation of the noise in lensing 
reconstruction. We also acknowledge using CAMB for power spectra calculations.
CK gratefully acknowledges support from DOE grant DE-FG03-91ER40662 
and NASA ATFP grant NNX08AL48G.
BK gratefully acknowledges support from NSF PECASE Award AST-0548262.
GMF, CS and CK acknowledge partial support from NSF grant PHY-06-52626 
at UCSD.

\newpage

\begin{table}
\begin{centering}
\begin{tabular}{c c c c c c}
\hline
Experiment & $f_{\rm sky}$ & $\nu$ [GHz] & $\theta_b$ [1'] & 
$\Delta_T$ [$\mu$K] & $\Delta_E$ [$\mu$K]\\
\hline
\hline
{\sc PLANCK} & 0.65
    &  30 & 33  &  4.4 &  6.2\\
&   &  44 & 23  &  6.5 &  9.2\\
&   &  70 & 14  &  9.8 & 13.9\\
&   & 100 & 9.5 &  6.8 & 10.9\\
&   & 143 & 7.1 &  6.0 & 11.4\\
&   & 217 & 5.0 & 13.1 & 26.7\\
&   & 353 & 5.0 & 40.1 & 81.2\\
&   & 545 & 5.0 & 401  & $\infty$\\
&   & 857 & 5.0 & 18300 & $\infty$\\
\hline
{\sc POLARBEAR} & 0.03
  & 90 & 6.7 & 1.1 & 1.6\\
& & 150 & 4.0 & 1.7 & 2.4\\
& & 220 & 2.7 & 8.0 & 11.3\\
\hline
{\sc EPIC} & 0.65
    &  30 & 28  &  0.5 &  0.7\\
&   &  45 & 19  &  0.3 &  0.4\\
&   &  70 & 12  &  0.2 & 0.3\\
&   & 100 & 8.4 &  0.2 & 0.3\\
&   & 150 & 5.6 &  0.3 & 0.4\\
&   & 220 & 3.8 & 0.7 & 0.9\\
&   & 340 & 2.5 & 2.2 & 3.2\\
&   & 500 & 1.7 & 9.4  & 13.3\\
&   & 850 & 1.0 & 740 & 1047\\
\hline
\end{tabular}
\caption{
Sensitivity parameters of the CMB experiments 
considered in this work:
$f_{\rm sky}$ is the observed fraction of the sky,
$\nu$ is the center frequency of the channels in GHz,
$\theta_b$ is the full width at half maximum 
in arc-minutes, $\Delta_{T}$ is the temperature
sensitivity per pixel in $\mu$K and $\Delta_E=\Delta_B$ is the
polarization sensitivity. \label{tab:specs}}
\end{centering}
\end{table}

\end{document}